\documentclass[10pt,journal,compsoc]{IEEEtran}

\usepackage[T1]{fontenc}
\usepackage[numbers,sort&compress]{natbib}
\usepackage{amsmath,amssymb,amsfonts}
\usepackage{algorithmic}
\usepackage{graphicx}
\usepackage{url}
\usepackage{xurl}
\usepackage{textcomp}
\usepackage{xcolor}
\usepackage{listings} 
\usepackage{xspace}
\usepackage{comment}
\usepackage[switch]{lineno}
\usepackage{tikz}
\usetikzlibrary{arrows.meta,positioning,fit,calc,backgrounds}
\usepackage{tikzsymbols}
\usepackage[nogroupskip,nonumberlist,nopostdot]{glossaries}
\usepackage{multirow,booktabs,tabularx}
\usepackage[caption=false,font=footnotesize]{subfig}
\usepackage{float}  
\usepackage{enumitem}
\usepackage{pifont}
\usepackage{relsize}
\usepackage{xurl}
\usepackage{hyperref}
\usepackage{cleveref}

\makeatletter
\newcommand{\runinhead}[1]{%
  \@startsection{runinhead}{4}{\z@}%
    {1.5ex plus .5ex minus .2ex}
    {-0.6em}
    {\normalfont\normalsize\sffamily\itshape}%
    *{#1}}
\makeatother

\newcommand{\passatk}{$\mathrm{pass}@k$\xspace}
\newcommand{\passatone}{$\mathrm{pass}@1$\xspace}

\newcommand{\totalExecutedLanguages}{7\xspace}
\newcommand{\totalExecutedModels}{13\xspace}
\newcommand{\totalMeasuredSolutions}{9,455\xspace}   

\newcommand{\latency}{latency\xspace}
\newcommand{\Latency}{Latency\xspace}
\newcommand{\energy}{energy consumption\xspace}
\newcommand{\Energy}{Energy consumption\xspace}
\newcommand{\memory}{peak memory\xspace}
\newcommand{\Memory}{Peak memory\xspace}

\renewcommand{\cite}{\citep}

\newcommand{\leetcode}{LeetCode\xspace}
\newcommand{\leetcodeDataset}{LeetCodeDataset\xspace}

\newacronym[plural={PAPs}]{pap}{PAP}{Performance Anti-Pattern}

\newacronym{loc}{LoC}{Lines of Code}

\newacronym{ux}{UX}{User eXperience}

\newacronym[plural={SATs}]{sat}{SAT}{Static Analysis Tool}

\newacronym{samu}{SAMU}{Static Analysis Mining Utility}

\newacronym{jvm}{JVM}{Java Virtual Machine}

\newacronym{jit}{JIT}{Just-in-Time}
\newcommand{\jit}{\gls*{jit}\xspace}
\newacronym{dce}{DCE}{Dead-Code Elimination}
\newcommand{\dce}{\gls*{dce}\xspace}
\newacronym[plural={LLMs}]{llm}{LLM}{Large Language Model}
\newcommand{\llm}{\gls*{llm}\xspace}
\newcommand{\llms}{\glspl*{llm}\xspace}
\newacronym{clbg}{CLBG}{Computer Language Benchmarks Game}
\newcommand{\clbg}{\gls*{clbg}\xspace}
\newacronym[plural={ASTs}]{ast}{AST}{Abstract Syntax Tree}

\newacronym{ide}{IDE}{Integrated Development Environment}

\newacronym{rest}{REST}{Representational State Transfer}
\newcommand{\rest}{\gls*{rest}\xspace}
\newacronym[plural={APIs}]{api}{API}{Application Programming Interface}
\newcommand{\api}{\gls*{api}\xspace}
\newcommand{\apis}{\glspl*{api}\xspace}

\newacronym{rapl}{RAPL}{Running Average Power Limit}
\newcommand{\rapl}{\gls*{rapl}\xspace}
\newacronym{pca}{PCA}{Principal Component Analysis}
\newcommand{\pca}{\gls*{pca}\xspace}

\newacronym{llama}{Llama-3.3}{Llama-3.3-70B-Instruct-Turbo}

\newacronym{gpt}{GPT}{GPT-4.5}

\newacronym{geminitwo}{Gemini-2.0}{Gemini-2.0-flash}

\newacronym{geminitwodotfive}{Gemini-2.5-pro}{Gemini-2.5-pro}

\newacronym{gemini}{Gemini}{Gemini-2.5-pro}

\newacronym{Mistral}{Mistral}{mistral-nemo-instruct-2407}

\newcommand{\gptastra}{\code{gpt-6-astra}}
\newcommand{\gptsol}{\code{gpt-5.6-sol}}
\newcommand{\gptluna}{\code{gpt-5.6-luna}}
\newcommand{\gptfivefive}{\code{gpt-5.5}}
\newcommand{\gptfivemini}{\code{gpt-5.4-mini}}

\newcommand{\opusfive}{\code{claude-opus-5-5}}
\newcommand{\opusfour}{\code{claude-opus-4-8}}
\newcommand{\sonnet}{\code{claude-sonnet-5-5}}
\newcommand{\fable}{\code{claude-fable-5-1}}
\newcommand{\deepseek}{\code{deepseek-v4.1-flash}}
\newcommand{\geminiflash}{\code{gemini-3.8-flash}}
\newcommand{\grok}{\code{grok-4.6}}
\newcommand{\gemma}{\code{gemma4:e4b}}
\newcommand{\glm}{\code{glm-5.3-flash}}

\newcommand{\perfarena}{\textsc{PerfArena}\xspace}

\newcommand{\clbgComp}{\code{CLBGComp}}
\newcommand{\leetcodeComp}{\code{LeetCodeComp}}

\newcommand{\RQ}[1]{\textbf{RQ#1}\xspace}

\definecolor{rqbg}{HTML}{EEF2F7}
\definecolor{rqfr}{HTML}{34547A}

\newcommand{\roundbox}[1]{
	\begin{center}
		\begin{tikzpicture}
			\node[draw=black, rectangle, rounded corners, inner sep=2mm](box){
				\begin{minipage}{0.9\columnwidth}
					#1
				\end{minipage}
			};
		\end{tikzpicture}
	\end{center}
}

\DeclareRobustCommand*{\circnum}[1]{%
  \tikz[baseline=(char.base)]{
    \node[
      circle,
      draw=black,
      fill=black,
      text=white,
      line width=0.4pt,
      inner sep=0.2pt,
      outer sep=0pt,
      minimum size=0pt,   
      text width=         
    ] (char) {#1};
  }%
}

\newcommand{\bench}[1]{\code{#1}}

\definecolor{darkred}{rgb}{0.75,0,0}
\newcommand{\rr}[1]{#1}

\newcommand{\kasolved}[1]{\textcolor{orange}{$\blacktriangleright$KA:\ding{51}?$\blacktriangleleft$}}

\newcommand{\jpfsolved}[1]{\textcolor{purple}{$\blacktriangleright$JPF:\ding{51}?$\blacktriangleleft$}}
\newcommand{\rrsolved}[1]{\textcolor{blue}{$\blacktriangleright$RR:\ding{51}$\blacktriangleleft$}}

\newcommand{\code}[1]{{\smaller\texttt{#1}}\xspace}
\newcommand{\class}[1]{{\smaller\urlstyle{tt}\nolinkurl{#1}}\xspace}

\definecolor{sh_comment}{rgb}{0.12, 0.38, 0.18}
\definecolor{sh_keyword}{rgb}{0.37, 0.08, 0.25}  
\definecolor{sh_string}{rgb}{0.06, 0.10, 0.98} 
\definecolor{KWColor}{rgb}{0.5,0,0.67}
\definecolor{CommentColor}{rgb}{0.15,0.5,0.15}
\definecolor{lightgrey}{rgb}{0.8,0.8,0.8}
\lstdefinelanguage{swift} {
  morekeywords = {open,catch,@escaping,nil,throws,func,if,then,else,for,in,while,do,switch,case,default,where,break,continue,fallthrough,return,typealias,struct,class,enum,protocol,var,func,let,get,set,willSet,didSet,inout,init,deinit,extension,subscript,prefix,operator,infix,postfix,precedence,associativity,left,right,none,convenience,dynamic,final,lazy,mutating,nonmutating,optional,override,required,static,unowned,safe,weak,internal,private,public,is,as,self,unsafe,dynamicType,true,false,nil,Type,Protocol},
  morecomment = [l]{//}, 
  morecomment = [s]{/*}{*/}, 
  morestring = [b]", 
}

\lstdefinestyle{Maple}{
	showspaces=false,showtabs=false,tabsize=2,columns=flexible,
	framesep=4pt,	
	showstringspaces=false,
	backgroundcolor=\color{white},
	breaklines=true,
	breakatwhitespace=false,
	xleftmargin=10pt,xrightmargin=0pt,
	basicstyle=\small\ttfamily,
	stringstyle=\color{sh_string},
	keywordstyle = \color{sh_keyword}\bfseries,
	commentstyle=\color{sh_comment}\itshape,
	numberstyle=\footnotesize\color{CommentColor},
	numbers=left,firstnumber=last,stepnumber=1,
	lineskip=-0.0em,
	escapeinside={/*@}{@*/},
	numbersep=5pt,
	captionpos=b,
	upquote=true,
}

\begin{document}

\title{PerfArena: Ranking LLMs by Code Performance}

\author{Rui~Rua, Hailemariam~Mersha, Jo\~{a}o~Paulo~Fernandes, and Karim~Ali%
\IEEEcompsocitemizethanks{\IEEEcompsocthanksitem R.~Rua, H.~Mersha, and K.~Ali are with New York University Abu Dhabi, Abu Dhabi, UAE.\protect\\
E-mail: rui.rua@nyu.edu, karim.ali@nyu.edu
\IEEEcompsocthanksitem J.~P.~Fernandes is with Universidade do Algarve, Faro, Portugal.\protect\\
E-mail: jpsfernandes@ualg.pt}}

\IEEEtitleabstractindextext{%
\begin{abstract}

Although \llm-generated production code is mostly correct, its fitness for repeated and extensive use depends on what it costs in time, memory, and energy. Nonetheless, most code-generation benchmarks evaluate correctness in isolation, and most performance-oriented benchmarks cover few languages and rarely measure \energy. We argue that evaluating code-generation \llms requires reproducible measurement of what their correct code costs, per task and language. To meet this requirement, we present \perfarena, an open-source arena that ranks \llms by the \latency, \memory, and \energy of their correct code. \perfarena measures every correct solution under one protocol, comparing two solutions only within one problem in one language. We evaluate \perfarena on two competitions: \clbgComp (10~Computer Language Benchmarks Game programs) and \leetcodeComp (99~\leetcode problems with measurable \latency differences). Across \totalMeasuredSolutions correct solutions from \totalExecutedModels~\llms in \totalExecutedLanguages~languages, \gptastra is the only \llm that no other \llm beats significantly in \latency, \energy, or \memory. The correct solutions of the other \llms take 1.1--1.6$\times$ as long as those of \gptastra on \leetcodeComp and 1.2--3.6$\times$ on \clbgComp. The most expensive \llms per correct solution cost 2.6$\times$ (\leetcodeComp) and 9.3$\times$ (\clbgComp) as much as \gptastra, yet do not generate faster code. \perfarena leaderboards and solutions are public and continuously updated, helping developers choose an \llm per language and metric rather than by pass rate or price.

\end{abstract}


}

\maketitle
\IEEEdisplaynontitleabstractindextext
\IEEEpeerreviewmaketitle

\section{Introduction}
\label{sec:intro}

\glsresetall

\llms now generate a substantial and growing share of the world's code. For example, Google reports that the share of its new AI-generated code grew from more than 25\% in late 2024~\cite{Pichai2024Q3} to 75\% in 2026~\cite{Pichai2026Blog}. As developers delegate more implementation work to \llms, the efficiency of the shipped software increasingly depends on the efficiency of the code that \llms produce. However, the community still judges \llm-generated code mostly by its functional correctness. For example, HumanEval~\cite{Chen2021HumanEval} counts a program as a success if it passes its unit tests (\passatk). Later benchmarks, such as EvalPlus~\cite{Liu2023EvalPlus} and LiveCodeBench~\cite{Jain2025LiveCodeBench}, strengthen these tests or refresh the problems without changing this objective. Two factors compound this correctness focus. First, \llm developers optimize and report progress mostly on functional correctness. Second, \llms learn from human-written code, which rarely prioritizes performance. These factors amplify the performance inefficiencies already present in human-written code. \citet{rua_ali_paps_conf} call this feedback loop a \emph{performance loop of hell}, estimating a 10$\times$ density of performance anti-patterns in \llm-generated Android apps relative to human-written apps. Other studies report that \llms struggle to analyze~\cite{nofuneval_microsoft} and optimize~\cite{gso_challenging_optimization,tom_ana_llms_python} software performance.

Existing efficiency benchmarks each cover only part of this picture. EffiBench~\cite{EffiBench} measures \latency and \memory. Mercury~\cite{Mercury} and ENAMEL~\cite{ENAMEL} primarily evaluate \latency, while EvalPerf~\cite{EvalPerf} counts executed CPU instructions. Most evaluate a single language, and the broadest, EffiBench-X~\cite{Qing2025EffiBenchX}, covers 6~languages without measuring \energy. EnergyBench~\cite{Ionescu2026EnergyBench} combines functional correctness with energy-efficiency measurements but studies correctness-energy trade-offs and prompt sensitivity rather than ranking \llms. Several empirical studies~\cite{Islam2025Energy,tom_ana_llms_python,Solovyeva2025Forge} also measure the \energy of \llm-generated code, each with its own task set, hardware platform, and measurement protocol. These differences make their results hard to reproduce, and a new \llm needs a fresh evaluation under each setup. Therefore, efficiency evaluation lacks what correctness evaluation already has: an arena that measures the \latency, \memory, and \energy of every correct solution under one protocol, ranking an \llm against others without repeating the study.

To bridge this gap, we present \perfarena, an arena that ranks \llms by the \latency, \memory, and \energy of their correct programs. \perfarena covers more languages than prior efficiency benchmarks, namely \totalExecutedLanguages of the 8~most used languages on GitHub in 2025~\cite{GitHub2025Octoverse}. To organize its evaluation, \perfarena uses competitions, each a named set of coding problems with fixed specifications. It runs every solution through a correctness oracle and measures the correct ones with a performance oracle under the same protocol in every language. Since \perfarena compares two solutions only when they solve the same problem in the same language, its rankings never confuse the efficiency of an \llm with that of a language. We evaluate \perfarena on two competitions: \clbgComp, 10~compute-intensive benchmarks from the \clbg~\cite{clbg}, and \leetcodeComp, 99~\leetcode~\cite{leetcode} problems whose correct solutions differ measurably in \latency. We prompt \totalExecutedModels~\llms with one prompt template and collect their solutions in C++, C\#, Java, JavaScript, PHP, Python, and TypeScript.

Our results show that correct code is not necessarily efficient code. Although 12~\llms pass more than 90\% of their \leetcodeComp attempts, the correct solutions of the slowest \llm take on average 1.6$\times$ as long to run as those of the fastest (\gptastra). On \clbgComp, this gap grows to 3.6$\times$. Across both competitions, \gptastra is the only \llm that no other \llm beats significantly in \latency, \energy, or \memory. However, the fastest \llm depends on the language. For example, on \leetcodeComp, \opusfive leads in Python and PHP, and \gptastra in the other five languages. Moreover, beyond avoiding the cheapest \llms on \clbgComp, paying more does not buy faster code. Although \gptastra has one of the highest prices per token, it costs less than half as much per correct solution as the two most expensive \llms on \leetcodeComp (\fable and \gptfivefive), and neither generates faster code. The leaderboard, with every solution and its measurements, is live at \href{https://perfarena.net}{\nolinkurl{perfarena.net}}, and the source code is at \href{https://github.com/perfarena/perfarena}{\nolinkurl{github.com/perfarena/perfarena}}.

\section{Overview of \perfarena} \label{sec:perfarena}

\perfarena organizes evaluation into \emph{competitions}, each a named set of coding \emph{problems} with fixed specifications. To enter a competition, a source (i.e., an \llm, an AI workflow, or an agentic system) bundles its candidate solutions for one or more problems of the competition in one language, together with the \llm, prompt, and configuration that generated them, into a \emph{submission}. \Cref{fig:workflow} shows the \perfarena workflow: competition and problem selection~(\circnum{1}), code generation~(\circnum{2}--\circnum{3}), correctness validation~(\circnum{4}--\circnum{5}), execution measurement~(\circnum{6}--\circnum{7}), and results analysis~(\circnum{8}).

\perfarena enforces two constraints. First, correctness validation strictly precedes measurement, preventing incorrect programs from claiming efficiency gains. Second, \perfarena compares solutions only within a \emph{cell} (i.e., one problem in one language), holding task, workload, toolchain, and execution environment constant. To achieve that, \perfarena uses two interfaces: a \emph{correctness oracle} that assigns a verdict to a program, and a \emph{performance oracle} that measures \latency, \memory, and \energy. \perfarena implements both interfaces through the \leetcode judge (via its submission \api) or through custom harnesses that run in isolated environments. Every submission names exactly one correctness oracle and one performance oracle, tying each measurement to a specific judging backend and execution environment. \perfarena also records the provenance of every result, from prompts and decoding parameters to \llm digests, execution specifications, and profiling tools.

\perfarena exposes its pipeline through a \rest \api. A \emph{campaign} runs one or more competitions over a chosen set of \llms under a fixed measurement profile. Template client scripts drive a campaign under a chosen interaction workflow (e.g., single-shot generation). The web app reads the same \api to let users browse competitions, leaderboards, and the provenance of any submission. Instead of running programs itself, the \api records each submission, resolves its two oracles, and dispatches the work asynchronously. Hence, a client polls for the verdict instead of blocking on a slow judge. \perfarena stores one measurement row per run, stamped with the execution environment, the energy tool, and a unique run identifier. Therefore, every number on a leaderboard traces back to the hardware and the tool that produced it.

\begin{figure}
\centering
\resizebox{\columnwidth}{!}{%
\begin{tikzpicture}[
  font=\scriptsize,
  box/.style={rounded corners=2.5pt, draw, align=center, inner sep=2pt, font=\scriptsize, text width=19mm, minimum height=6.5mm},
  usr/.style={box, minimum height=5mm},
  core/.style={box, text width=14mm, minimum height=6mm},
  iface/.style={core, text width=8mm, minimum height=17.5mm},
  srv/.style={box, text width=22mm, minimum height=5.5mm},
  note/.style={font=\tiny\itshape, text=black!55, align=center, inner sep=1pt},
  cont/.style={rounded corners=4pt, draw=black!60, line width=.7pt},
  ctitle/.style={font=\scriptsize\bfseries, anchor=south, inner sep=1.2pt},
  both/.style={{Latex[length=1.3mm]}-{Latex[length=1.3mm]}, line width=.6pt},
  iarr/.style={-{Latex[length=1.1mm]}, line width=.5pt},
  lbl/.style={font=\scriptsize, text=black!70, align=center, inner sep=1pt},
]
\node[box, anchor=west] (llm) at (0,-0.6) {\textbf{LLM}\\OpenAI API};
\node[box, anchor=west] (cli) at (0,-1.6) {\textbf{Client}\\client scripts};
\node[usr, anchor=west] (web) at (0,-2.35) {\textbf{Web App}};
\node[usr, anchor=west] (usr) at (0,-3.2) {\textbf{User}};
\node[iface] (api)  at (4.5,-1.975) {REST\\API};
\node[core]  (disp) at (6.55,-1.3) {Dispatcher};
\node[core]  (db)   at (6.55,-2.65) {Database};
\begin{scope}[on background layer]
  \node[cont, fit=(api)(disp)(db), inner sep=1.8mm] (be) {};
\end{scope}
\node[srv]  (csrv) at ($(be.south)+(-2.15,-1.6)$) {HTTP Server \texttt{/validate}};
\node[note, below=0.4mm of csrv, text width=30mm] (cwork) {\mbox{runs the validation workload}};
\node[srv]  (psrv) at ($(be.south)+(2.15,-1.6)$) {HTTP Server \texttt{/measure}};
\node[note, below=0.4mm of psrv, text width=30mm] (pwork) {\mbox{runs the measurement workload}};
\begin{scope}[on background layer]
  \node[cont, fit=(csrv)(cwork), inner sep=1.3mm] (cor) {};
  \node[cont, fit=(psrv)(pwork), inner sep=1.3mm] (per) {};
\end{scope}
\node[ctitle, anchor=south west] at (cor.north west) {Correctness Oracle};
\node[ctitle, anchor=south east] at (per.north east) {Performance Oracle};
\node[ctitle] at (be.north) {Backend};
\coordinate (bl) at ($(be.south west)!0.3!(be.south east)$);
\coordinate (br) at ($(be.south west)!0.7!(be.south east)$);
\draw[iarr] (api.east |- disp) -- (disp.west);
\draw[iarr] (disp) -- (db);
\draw[iarr] (db.west) -- (api.east |- db);
\draw[both] (llm.south -| cli.north) -- (cli.north) node[lbl,pos=.5,left=1mm]{\circnum{2}\,prompt / code};
\draw[both] (cli.east) -- (be.west |- cli) node[lbl,align=left,pos=.5,above=0.2mm]{\circnum{1}\,fetch\\\circnum{3}\,submit\\\circnum{8}\,results};
\draw[both] (web.east) -- (be.west |- web) node[lbl,pos=.5,above=0.2mm]{results} node[lbl,pos=.5,below=0.2mm]{leaderboard};
\draw[both] (web.south) -- (usr.north) node[lbl,pos=.5,right=1mm]{browse};
\draw[both] (bl) -- (bl |- cor.north) node[lbl,align=left,pos=.5,left=0.5mm]{\circnum{4}\,validate\\\circnum{5}\,verdict};
\draw[both] (br) -- (br |- per.north) node[lbl,pos=.5,left=0.5mm,font=\tiny\itshape,text=black!55]{only if\\accepted} node[lbl,align=left,pos=.5,right=0.5mm]{\circnum{6}\,measure\\\circnum{7}\,metrics};
\end{tikzpicture}%
}
\caption{The \perfarena workflow.}
\label{fig:workflow}
\end{figure}

\section{Building Competitions} \label{sec:benchmark-selection}

\perfarena currently hosts two competitions, \clbgComp and \leetcodeComp, each built from a different problem source. We source \clbgComp from the 10~whole programs of the \clbg~\cite{clbg}, widely used in prior software performance studies~\cite{Pereira2017Energy,scp21,easy_green_ase25,cross_platform_wellington}. For \leetcodeComp, we sample 99~problems from a corpus of 1,973~public \leetcode problems~\cite{leetcode,LimYeri2025LeetCodeSolutions}, keeping those whose \latency differences best separate an efficient solution from an inefficient one. Every problem carries its specification (i.e., description, inputs, arguments, data types, and examples) for each programming language. We take each problem statement and its reference solutions from this corpus. We then retrieve its difficulty, its problem categories, and its per-language starter code through \leetcode's GraphQL \api~\cite{graphql}. To evaluate each problem, \perfarena uses two distinct test suites: a comprehensive \emph{validation workload} to establish functional correctness, and a curated \emph{measurement workload} for performance analysis.

\subsection{Problem Selection} \label{sec:problem-selection}
Since \clbgComp problems are computationally intensive by design, they naturally distinguish efficient from inefficient solutions. In contrast, many \leetcode problems are computationally trivial (e.g., deciding whether a given word is a palindrome). For these problems, the execution times of all accepted solutions are practically identical, and their variance is often smaller than the measurement noise or the monitoring overhead itself. Therefore, we select the \leetcode problems that exhibit measurable and distinguishable performance differences, using the historical \latency of the accepted submissions on the \leetcode platform.

First, we collect \latency distributions. For each candidate problem, we submit a Python probe solution from the problem-and-solution corpus~\cite{LimYeri2025LeetCodeSolutions} to unlock \leetcode's statistics endpoint. We then query the GraphQL \api for the \latency distribution of accepted Python submissions, which \leetcode builds from the runtimes of its users and returns as a histogram of \emph{buckets} (i.e., discrete time intervals). From this distribution, we compute the 10\textsuperscript{th} (fast), 50\textsuperscript{th} (typical), and 95\textsuperscript{th} (slow) percentiles. Second, we filter measurable problems. We retain a problem only if (1)~its median \latency is at least 10~ms, removing problems that run too fast to measure reliably; (2)~95\textsuperscript{th}--10\textsuperscript{th} percentiles gap is at least 10~ms, removing problems where fast and slow solutions barely differ; and (3)~its distribution spans at least 15~buckets, removing problems whose latencies fall into so few buckets that their percentiles are coarse and unstable. Third, we rank the remaining problems by the ratio of their 95\textsuperscript{th} to their 10\textsuperscript{th} percentile, which tells how many times slower a slow submission is than a fast one. A high ratio marks a problem where the choice of algorithm strongly changes \latency. The ratio also cancels any factor that scales all latencies alike, such as the number of test cases that a problem runs.

Overall, this procedure distills the 1,973~candidate problems down to the 99~problems of \leetcodeComp. After we drop the problems that the \leetcode judge cannot time reliably, 886~candidates clear the thresholds. Removing the 22~\leetcode Premium-only problems leaves 864 (151~Easy, 486~Medium, and 227~Hard), from which we keep the 33~problems with the highest ratio per tier.

\subsection{Validation Workload} \label{sec:validation-workload}
\perfarena compares the output of \clbgComp programs against reference outputs that the C++ reference programs produce at small arguments. For text outputs (e.g., the DNA sequences that \bench{fasta} prints), the comparison ignores differences in white space and case. \perfarena also compares the numbers that \bench{n-body} and \bench{spectral-norm} print within an absolute tolerance of 10$^{-6}$. For the binary bitmap of \bench{mandelbrot}, \perfarena allows at most 16~differing bytes, because floating-point rounding at the boundary of the set differs slightly between languages.

\perfarena draws the validation test cases of \leetcodeComp from \leetcodeDataset~\cite{Xia2025LeetCodeDatasetHF}. We drop cases whose inputs fall outside a problem's stated bounds or are otherwise malformed, leaving 9,188~test cases over the 99~problems. At run time, \perfarena runs each solution on the shared, language-independent inputs and compares its output, in a canonical form, against the reference. Each language has a harness that reads the same file of inputs and expected outputs of a problem, enabling \perfarena to validate a solution in any of its languages on all cases without the \leetcode judge. Since several outputs may be correct for some problems, \perfarena applies an \emph{order-insensitive validator} that accepts legitimate alternative orderings (e.g., sets or graph paths). In our campaign, the \leetcode judge produces every \leetcodeComp verdict that we report.

\subsection{Measurement Workload} \label{sec:test-selection}

\paragraph*{\clbgComp}
Most \clbgComp benchmarks run as a whole program, each with one integer argument that sets the problem size. Only 3~benchmarks (\bench{k-nucleotide}, \bench{regex-redux}, and \bench{reverse-complement}) read their input differently, from the file that the C++ reference program of \bench{fasta} generates. On our measurement host, the canonical \clbg arguments (e.g., 21 for \bench{binary-trees}) make the slowest languages exceed the 1,200-second limit of one measurement job. Conversely, a small argument makes C++ finish within milliseconds, below the noise floor of the supported energy profilers: \rapl hardware counters~\cite{rapl} and CodeCarbon package-energy estimates~\cite{codecarbon}. Therefore, we calibrate the argument of each benchmark to the measurement host between two bounds. The lower bound comes from C++, the fastest of our measured languages~\cite{scp21}, which must run for about 1~second to rise above that noise floor. The upper bound comes from Python, which is slower than the other languages~\cite{Pereira2017Energy,scp21} and must still finish its repeated executions within the job limit.

The two bounds cannot both be tight, because the fastest and slowest languages may differ by more than an order of magnitude on one problem. Prior studies report such gaps~\cite{Pereira2017Energy}, as does our host (e.g., the median Python solution of \bench{binary-trees} runs $\approx$~51$\times$ longer than the median C++ one). To pick an argument between both bounds, we time the C++ and Python programs at candidate arguments and choose one at which C++ runs for about 1~second or more and Python still finishes within the job limit. Mapping that argument to every language keeps a cross-language comparison from confounding efficiency with workload size. Changing the measurement host requires recalibration, because a faster/slower environment shifts both bounds. We fix the calibrated arguments for our campaign, because measurements at different ones are not comparable.

\paragraph*{\leetcodeComp}
Measuring every one of the 9,188~test cases ($\approx$~93 per problem) across every problem, language, and \llm does not scale. To capture a reliable performance footprint, code must run long enough for measurement tools to gather multiple samples. However, test cases that trigger identical code paths (e.g., same branches) add overhead without new insight. Therefore, we condense each problem into a small set of test cases that still covers every distinct \emph{kind} of work that the code performs. We then reuse that set across all languages and \llms. Since every program must pass a full correctness check before measurement, this reduced set serves strictly as a measurement workload.

To compute the reduced set, we start from the test cases that \leetcodeDataset~\cite{Xia2025LeetCodeDatasetHF}, a public dataset of Python \leetcode problems, pairs with each \leetcodeComp problem. We then cap each problem at 100~test cases, sampling randomly under a fixed seed. To select the measurement subset, we adapt a fingerprinting and clustering procedure from SimPoint~\cite{Sherwood2002SimPoint}. On a bare-metal Intel Linux host that exposes \code{perf}~\cite{perf} counters and \rapl~\cite{rapl} energy readings, we run each sampled test case in a loop for 0.3~seconds and record a behavioral fingerprint from six hardware counters: retired instructions, cycles, branches, branch misses, cache references, and cache misses. We also record the \energy of each case to identify the highest-energy case. The cap leaves 8,103~cases, and we profile 8,063 of them, because 40~cases across 5~problems fail in the harness.

The fingerprint of a test case is a vector of the logarithms of its counter values per call, together with its cache-miss rate, branch-miss rate, and instructions per cycle. We use these fingerprints to reduce each problem's cases to a small representative set. First, we rescale each counter to zero mean and unit variance, preventing a large-magnitude counter (e.g., instruction count) from dominating a smaller one (e.g., branch misses). We then apply \pca and keep enough components to preserve 90\% of the variation, turning several correlated counters into a few independent axes. On those axes, we cluster each problem's cases with $k$-means, choosing the number of clusters by the best silhouette score, a standard measure of how cleanly clusters separate~\cite{Rousseeuw1987Silhouettes}. Since test cases in the same cluster do the same kind of work, we keep only the case nearest to each cluster centroid (i.e., the medoid). We also keep the single highest-energy case, because an inefficient solution shows its true cost on the heaviest input, which clustering toward the average might discard. If a problem has fewer than 5~kept cases, we add the case least similar to those already kept (i.e., a maximin choice), one at a time, until the set reaches~5. Across all \leetcodeComp problems, this procedure keeps 501~profiled cases (6.2\%), 5~per problem at the median and at most~8.

To keep the average over the kept cases representative of the whole problem, \perfarena assigns every original case to the kept case that it most resembles. The weight of a kept case, including the highest-energy case, is the fraction of the problem's cases assigned to it, such that all weights of a problem add up to~1. This weighted average approximates the average over all cases at a fraction of the measurement time. The kept cases may shift the values of a problem. However, \perfarena compares a solution only with the other solutions of its cell, where a shift that all solutions share cancels. Only a change in the order of two solutions matters, and such changes are rare. On 20~cells that we measure on all their profiled test cases, the kept cases reverse which of two solutions is faster for 1.5\% of the pairs, and which uses less \energy for 0.6\%. We opt for this behavior-based set compared to simply the largest inputs, because input size alone does not capture the behavior of a test case. Input size correlates with the main fingerprint axis at a median absolute correlation of 0.94 over the problems (mean 0.82), accounting for much of that behavior but not all of it. We record the random seeds, the fingerprints, and the kept cases, enabling others to reproduce the exact selection.

\subsection{Translating \leetcodeComp Test Cases} \label{sec:translation}
Every language-specific implementation of a \leetcodeComp problem reads identical test-case inputs from the same JSON files. Hence, we reuse the minimized Python test suites for the other languages, without translating the inputs themselves. From each problem's canonical function signature (i.e., method name, parameter types, and return type), we generate a standalone module per language with a solution stub, a harness that parses the shared input and runs the function, and a build script that standardizes compilation and execution. The harness handles simple types (e.g., scalars, arrays) directly and uses dedicated parsers for complex structures (e.g., trees, linked lists, and stateful objects). Each implementation folds every result into a checksum to prevent \dce by a compiler.

To translate from Python, we use Claude Opus~4.8~\cite{anthropic_claudeopus48}, the leading \llm when we started developing \perfarena. The same \llm generates the harness code (i.e., parsers for complex structures, the runner, and the checksum), the test code of each case whose input needs a language-specific construction (e.g., the operation sequence of a stateful design problem), and a reference solution per language. The \llm also generates the tests one case at a time to localize translation errors to single cases. We validate each generated case against the \leetcode judge, using correct solutions~\cite{LimYeri2025LeetCodeSolutions} and failing solutions from \code{qwen2.5:7b} and \gemma, two older \llms that we exclude from the campaign. When a generated case produces a verdict that disagrees with the \leetcode judge, we regenerate it until all verdicts agree. The reference solution of each language must also pass every case of the validation workload. The validated cases form the final measurement workload.

\section{Generating and Validating Solutions}
\label{sec:submission}

For each competition, \perfarena first obtains candidate programs from every \llm (\circnum{2}--\circnum{3}), and then admits only correct programs to measurement (\circnum{4}--\circnum{5}).

\subsection{Code Generation}
\perfarena supports any \llm served through an OpenAI-compatible \rest \api~\cite{openai_api}, a standard interface that covers the major \llm families such as Claude, Gemini, and GPT. To construct a query, \perfarena uses a pre-defined prompt template that provides the problem specification, the required language, the input and output contract, and any language-specific compilation or interface rules. Every \llm receives the same prompt for a given problem, and only the language-specific part changes across languages. Within each cell, \perfarena fixes the prompt version and the decoding configuration, drawing the same number of independent samples from each \llm (\circnum{2}--\circnum{3}).

The query generation stage is pluggable. Beyond the single-shot workflow in \Cref{fig:workflow}, \perfarena supports multi-step workflows, including a feedback-driven agent that plans, generates, and revises a solution before committing a final one. All submissions of our campaign use the single-shot workflow. Hence, our rankings compare \llms under the same workflow. \perfarena records the full generation trace in every case, and ships template client scripts for these workflows, starting with the single-shot one. A client script stores the full response and then extracts the code. \llms usually return their code inside a Markdown code block, which may carry a language tag (e.g., \code{python}). The script takes the first code block that carries the tag of the requested language, otherwise the first code block of any kind. If not code block exists, it takes the whole response. The script does not change the logic of the extracted program. Before a correctness oracle runs a program, it applies only mechanical fixes that its harness needs (e.g., standard Java imports and the PHP open tag on \clbgComp). \perfarena records programs that still fail to compile or run as failures and never re-runs them.

Each sample (i.e., an \llm response to a prompt) yields 3~files: the extracted source, the raw response, and a metadata record. The metadata records the provider and \llm, the decoding parameters, the sample and problem identifiers, the language, the prompt, the raw response and any reasoning trace, a timestamp, the generation time, and a hash of this record. \perfarena carries these identifiers into the validation and measurement records. This trail keeps \llm updates, prompt revisions, and environment drift from passing as unexplained measurement noise.

\subsection{Correctness Validation}
To avoid wasting compute and polluting performance results with invalid solutions, \perfarena requires every program to pass a correctness oracle before it advances to measurement. For \leetcodeComp, \perfarena supports two correctness oracles: an external oracle that delegates judging to \leetcode's rate-limited online judge, and a custom offline oracle that runs the validation workload. Across both competitions, a correctness oracle assigns one of the following verdicts: \code{Accepted}, \code{Wrong Answer}, \code{Runtime Error}, \code{Time Limit Exceeded}, \code{Compile Error}, or \code{Memory Limit Exceeded}. A response in the wrong language, or one that the runner cannot dispatch, also counts as a failure. Only an \code{Accepted} program moves forward to the execution measurement pipeline (\circnum{6}--\circnum{7}).

Building compiled languages must not count toward \latency. Therefore, \perfarena compiles each program before the measurement campaign and outside all timed regions, using a fixed toolchain and optimization policy recorded in the campaign's measurement profile. Any change to the compiler settings creates a distinct profile, and \perfarena never pools measurements across profiles. \perfarena logs compilation failures separately from execution failures to distinguish language-generation errors from runtime defects. \perfarena still stores the verdict and code of every rejected program. Rather than charging failing code an arbitrary \latency or \energy penalty, \perfarena counts a rejected program as a miss in \passatk. This design decouples the ability of an \llm to produce functional code from the execution efficiency of that code.

As \perfarena fills its leaderboards, we audit solutions that a correctness oracle rejects. For each rejected solution, we reproduce the verdict and read the output difference to confirm that the rejection reflects a defect in the submitted code and not a limitation of the comparison. This audit extends the \clbgComp comparison beyond an exact match, because a correct whole-program solution may differ from the reference output in ways that the benchmark does not define. In particular, \perfarena (1)~normalizes the formatting that the benchmark leaves unspecified, collapsing repeated blanks, trimming each line end, and ignoring blank separator lines and a missing final newline; (2)~folds letter case for benchmarks whose reference output is case-free (e.g., \bench{reverse-complement}); (3)~relaxes line order for benchmarks whose output lines are independent and self-labeled, because a parallel solution may emit these lines in a different order (e.g., \bench{binary-trees}); and (4)~compares floating-point results under a per-problem numeric tolerance, which keeps a last-digit difference from summation order from rejecting a correct value (e.g., \bench{n-body} and \bench{spectral-norm}).

\section{Measuring and Ranking Efficiency} \label{sec:ranking}

\perfarena measures every accepted program (\circnum{6}--\circnum{7}) and aggregates the measurements into per-competition leaderboards (\circnum{8}).

\subsection{Execution Measurement}
The \leetcode judge reports \latency and \memory, but no \energy. Deriving our rankings from that oracle would give rapid results on an industry-standard platform without a time-consuming local measurement campaign. However, the \latency that the judge reports is inconsistent across the 5~submissions that we make for each program. Across the programs whose solutions we submit to the judge, the ratio between a program's largest and smallest \latency reading has a median of 1.26 (mean 1.29) and reaches 8.5~in the worst case. For \memory, the ratio is much smaller (median 1.05, at most 1.07). Therefore, to compute the \perfarena leaderboards using consistent measurements, we build a custom performance oracle that reports \latency, \memory, and \energy.

To provide reliable and fine-grained measurements, the \perfarena custom performance oracle executes validated programs on trusted, registered environments (\circnum{6}--\circnum{7}). The oracle may run on any Unix-like host and adopts the isolated measurement methodology established for prior studies on cross-language \energy~\cite{Pereira2017Energy,scp21}. The oracle runs each program in two passes: one measures \latency and \energy, and the other measures \memory. To measure \memory, \perfarena reads the maximum resident set size that the kernel records for the process (\code{ru\_maxrss}), the same value that GNU \code{time}~\cite{gnu_time} reports. Measuring \memory in a separate pass keeps the memory accounting from perturbing the \latency and \energy profiling. The oracle supports multiple profiling backends: \rapl~\cite{rapl}, CodeCarbon~\cite{codecarbon}, and powermetrics~\cite{powermetrics}. It defaults to the on-chip \rapl energy counters when the host exposes them, and otherwise falls back to software power-model estimates. To prevent external interference, \perfarena evaluates code on dedicated benchmarking hosts that run no \llm generation workloads. The oracle runs as a persistent local HTTP server that processes jobs sequentially, keeping its own startup, network overhead, and scheduling out of the reported numbers.

For \leetcodeComp, a single call of a solution on a small test case may last only microseconds, far shorter than the surrounding process startup and compilation, whose cost differs sharply between interpreted and compiled languages. Therefore, \perfarena runs each test case of the measurement workload under a shared \emph{per-invocation} harness. The harness first calls the solution repeatedly on the same input for 30\% of a wall-time budget (1.5~s by default), without counting these warmup calls, to let caches and \jit compilers settle. The harness then counts the calls that it completes in the remaining 70\% of the budget, and records the time of this measured window. We evaluate the suitability of the warmup share by measuring the \leetcodeComp solutions of \gptastra for 10~random problems in each language with the default budget and with a budget three times larger. With the longer warmup, the \latency of one call changes by a median of 0.8\%, a negligible shift in every language except C\#, where it falls by a median of 3.1\%. The energy tool (e.g., CodeCarbon) wraps the whole harness process. Hence, \perfarena assigns to the window the share of the total \energy that matches the share of the \latency of the window. Dividing the time and the \energy of the window by the number of calls gives the \latency and the \energy of one call, which we call the \emph{per-invocation} \latency and \energy. To compute one value per problem, \perfarena then takes the weighted mean over the test cases using the cluster weights of the measurement workload. 

\Latency excludes startup and compilation, whereas \energy excludes them only when it accrues at a roughly constant rate across the phases. \Memory of a solution is the largest peak over its test cases and covers the whole harness process, including the language runtime. A measurement fails when the program crashes or times out on a test case, or when the energy reading of the test case is missing. A failed measurement never counts as zero. Instead, \perfarena leaves that test case out of the weighted mean of every metric that it lacks and renormalizes the weights of the remaining test cases, reporting the measurement when at least one test case succeeds.

Besides the checksum that keeps a compiler or \jit from eliminating calls whose results go unused, the C++ harness passes every argument through an optimization barrier before each call. This barrier prevents the compiler from computing a repeated call on an unchanged input once and reusing its result. The harnesses for Python, JavaScript, TypeScript, PHP, Java, and C\# check once whether the solution changes its input, and give each call a fresh copy only if it does. The C++ harness copies the input for every call, adding the same cost to the \latency of every solution of a test case. For a stateful design problem, every harness rebuilds the object before each replay of the operation sequence.

For \clbgComp benchmarks, \perfarena executes the whole program once and discards this warmup run, bringing the input and the program files into the file cache of the operating system. \perfarena then executes a configurable number of measured runs, recording \latency, \energy, and the exit status of each run. A separate run measures \memory once per solution. Unlike \leetcodeComp, every run starts a new process without internal warmup. Hence, the \latency and \energy of a run include the startup of the language runtime.

For both competitions, \perfarena stamps every run with its measurement profile. A measurement profile records the machine (its processor, core count, and kernel version), the energy tool (e.g., \rapl counters or a CodeCarbon estimate), and the measurement configuration. When the runs of one submission report different energy tools or machines, \perfarena logs the conflict and leaves the energy tool of the submission unset. The leaderboard shows the machine and the energy tool of every row.

\subsection{Leaderboard Construction}
To rank \llms, \perfarena compares the accepted solutions of every two \llms on each cell that both solve, using a single sample that each \llm generates per cell, and fits a Bradley-Terry model to these \emph{duels}. To quantify the gap between two \llms, \perfarena divides each value by the value of the reference solution of its cell (i.e., the best correct solution that any \llm in the campaign produces for that cell and metric), and compares these ratios over the cells that both \llms solve. A cell in which an \llm has no accepted solution gives that \llm no duel, and \passatone reports these misses separately. The online leaderboard ranks \llms using the same approach as our evaluation and also shows two or more \llms side by side, with their results on every problem of a chosen language. For submissions that the external \leetcode oracle judges, \perfarena additionally records \latency and \memory percentiles that the judge reports against the other accepted submissions on the platform.

Leaderboards update as new submissions arrive, keeping the rankings current instead of frozen at a single point in time~\cite{Balunovic2025MathArena}. All leaderboards and their provenance records are public, but only we can add results to them, because we generate and measure every solution ourselves. Other researchers may deploy the open-source \perfarena on their own hosts and run their own leaderboards. Each ranking uses the measurements of one performance oracle, which may pool several machines with the same hardware and operating system. Before adding a machine to the pool, an administrator checks its hardware and operating system by hand. \perfarena checks that the median \latency and \energy of a fixed calibration solution on the new machine fall within 5\% of those on the first machine. Since we register a host with other hardware as a separate oracle, no ranking mixes hardware. In our campaign, one host measures every solution.

\section{Experimental Setup}
\label{sec:evaluation}

\begin{table}[t]
\centering
\footnotesize
\setlength{\tabcolsep}{4pt}
\caption{Selected \llms, their price in USD per 1M~tokens, and whether they sample at temperature~0 (\ding{51})  or at their default temperature (\ding{55}).}
\label{tab:models}
\resizebox{\columnwidth}{!}{
\begin{tabular}{llrc}
\toprule
\textbf{\llm} & \textbf{Provider} & \textbf{\$ per 1M (in/out)} & \textbf{Temp.~0} \\
\midrule
\gptastra      & OpenAI    & 5 / 25  & \ding{55} \\
\gptsol        & OpenAI    & 4 / 20  & \ding{55} \\
\gptluna       & OpenAI    & 0.2 / 1.2   & \ding{55} \\
\gptfivefive   & OpenAI    & 5 / 30  & \ding{55} \\
\gptfivemini   & OpenAI    & 0.37 / 2.25   & \ding{55} \\
\opusfive      & Anthropic & 4 / 20  & \ding{55} \\
\opusfour      & Anthropic & 5 / 25  & \ding{55} \\
\sonnet        & Anthropic & 2 / 10  & \ding{55} \\
\fable         & Anthropic & 10 / 50 & \ding{55} \\
\geminiflash   & Google    & 0.75 / 3.75   & \ding{55} \\
\deepseek      & DeepSeek  & 0.44 / 1.32   & \ding{55} \\
\grok          & xAI       & 2 / 6   & \ding{51} \\
\glm           & Zhipu     & 0.37 / 2.25   & \ding{51} \\
\bottomrule
\end{tabular}
}
\end{table}

\subsection{Selected \llms}
We evaluate \totalExecutedModels~\llms from 6~providers: OpenAI~\cite{openai}, Anthropic~\cite{anthropic}, Google~\cite{google_gemini38flash}, DeepSeek~\cite{deepseek_v41flash}, xAI~\cite{xai_grok46}, and Zhipu~\cite{zai_glm53flash}. The \llms range from flagship \llms to smaller, faster variants. \Cref{tab:models} lists all \llms that we evaluate across \totalExecutedLanguages~languages: TypeScript, Python, JavaScript, Java, C\#, PHP, and C++. These languages are \totalExecutedLanguages of the 8~most used languages on GitHub in 2025~\cite{GitHub2025Octoverse}. We leave out the remaining one, Shell, because \leetcode accepts Shell only for its shell-scripting problems and \clbg has no Shell programs.

\subsection{Prompt Template and Decoding}
To make the \llm the only variable within a cell, we give every \llm the same prompt template and draw one sample per cell. The template fixes the system role and the required response shape (e.g., function signature), easing code extraction. The supplemental material shows an example prompt. To generate solutions, we use a temperature of 0 for every \api that accepts one. However, the Gemini \api ignores this parameter for \geminiflash~\cite{api_gemini_38}, as does the DeepSeek \api in the default thinking mode of \deepseek~\cite{api_deepseek_thinking}. The OpenAI and Anthropic \apis accept only the default temperature for their reasoning \llms~\cite{api_openai_gpt6,api_claude_thinking}. Therefore, only \grok and \glm generate at temperature~0. Every other \llm uses the default of its \api. Since we draw one sample per cell, we do not measure how much an \llm's code varies between samples.

\subsection{Machine Configuration}
We measure on a single-purpose bare-metal host with an Intel Core i7-7600U processor (2~cores, 4~threads, 2.8~GHz base and 3.9~GHz Turbo Boost frequency) and 16~GB of RAM, which runs Ubuntu~22.04 and exposes \rapl. The host keeps its default frequency settings, with Turbo Boost enabled and the \code{powersave} governor of the \code{intel\_pstate} driver~\cite{intel_pstate}. To emulate a more realistic usage scenario, we do not fix the processor frequency, letting the driver adjust it to the load. We also do not pin a solution to a core.

We measure energy using CodeCarbon, which adds the package energy from \rapl counters to a RAM estimate (a constant 10~W on our host). We subtract this estimate to report only the package energy. The \rapl counters update about every millisecond~\cite{Khan2018RAPL}. Over the measured window (at least 1.05~s on \leetcodeComp), this interval adds at most 0.1\% error. We keep the idle power of the package (2--3\% of a reading), because subtracting it would change the \energy ratio of two solutions by at most 0.6\% on \leetcodeComp and 1.1\% on \clbgComp, below the tie margins. Of the 8 \clbgComp solutions whose \energy we discard because they run for less than 50~ms, 4~are \bench{binary-trees} solutions that skip the benchmark's work.

We run one measurement at a time, with the per-invocation harness on \leetcodeComp and with whole-program runs on \clbgComp. We repeat each \latency and \energy measurement 7~times. At the median coefficient of variation of a solution's runs (1.3\% on \leetcodeComp and 2.8\% on \clbgComp), 7~runs keep the 95\% confidence interval of the mean within $\pm$1.2\% and $\pm$2.6\%, respectively. For \energy, the median coefficients of variation are 2.8\% and 2.7\%, which keep the interval within $\pm$2.6\% and $\pm$2.5\%. For \memory, we measure once per solution, keeping the serial measurement time bounded. To check that one reading suffices, we run a pilot on our host with \gptastra on 20~random \leetcodeComp problems per language, taking 5~\memory readings of each solution. A single reading differs from the mean by a median of 0.1\% in TypeScript, 0.2\% in C\#, Java, and JavaScript, 0.3\% in PHP, 0.5\% in Python, and 1\% in C++. Only C\# and Java show larger outliers, with 95\% of their readings within 7\% and 5.7\% of the mean, respectively, against at most 2\% for the other languages.

\subsection{Computing Rankings} \label{sec:ranking-method}
To rank the \llms, we compare their solutions in pairs, similar to how Chatbot Arena compares chatbot answers~\cite{Chiang2024ChatbotArena}. Every cell that two \llms both solve gives one \emph{duel} per performance metric. The solution with the lower value wins that duel. Two values that differ by less than the measurement noise tie, giving half a win to each \llm. The noise of a difference between two measurements is $\sqrt{2}$ times the noise of one measurement. For \latency and \energy, the noise of one measurement is the 95\% confidence interval of its mean. Hence, two \latency values tie within 1.7\% on \leetcodeComp and 3.7\% on \clbgComp, and two \energy values tie within 3.7\% and 3.5\%. For \memory, the noise of one reading in a language is the 95th percentile of the deviation of a single reading from the mean in the pilot. Hence, two values tie within 0.7\% in TypeScript, 1\% in PHP, 1.5\% in JavaScript, 2.4\% in Python, 2.8\% in C++, 8.1\% in Java, and 9.8\% in C\#. We use the same margins on \clbgComp. A cell that an \llm fails gives it no duel. Therefore, a failure neither raises nor lowers its performance ranking, and \passatone reports these failures instead.

We fit a Bradley-Terry model~\cite{BradleyTerry1952} to the duels of each metric, using the MM algorithm~\cite{Hunter2004MM}. The model gives each \llm one strength. The chance that an \llm wins a duel is its strength divided by the sum of both strengths. From these strengths, we report the \emph{win probability} of each \llm, its mean chance of beating another \llm on a cell that both solve. To rank \llms within one language, we fit the model to the duels of that language only. An \llm with fewer than 3~solved problems in the language gets no rank there. With one or two solved problems, a single problem would decide at least half of the duels of the \llm in that language. This rule leaves out no \llm on \leetcodeComp and only 4/91~\llm-language pairs on \clbgComp (\gptfivemini in JavaScript and PHP, and \glm in C\# and PHP).

\subsection{Reference Set} \label{sec:reference-set}
For each problem, language, and performance metric, the \emph{reference set} holds the best correct solution of our campaign (completed 4 November 2026), which we call its \emph{reference solution}. A win probability tells how often an \llm wins, but not by how much. Therefore, we also report how far the code of each \llm is from the reference solution. The ratio of a solution divides its value by the value of the reference solution of its cell. A ratio of 1~marks the best solution, and 1.2~means 20\% worse. \Cref{tab:rq1-overall} reports the geometric mean of these ratios over the cells that an \llm solves as the ratio of that \llm. To compare two \llms, we divide the value of one \llm by that of the other in every cell that both solve. We then take the geometric mean of these quotients, following standard practice for normalized benchmark results~\cite{SPECCPU2026}. This comparison does not depend on the reference set. For example, a geometric mean of 1.2~means that the code of the first \llm takes 20\% longer to run than the code of the second. We freeze the reference set for future evaluations. Hence, adding an \llm that produces better solutions does not change the ratios of the \llms already evaluated.

In both competitions, every \llm produces at least one reference solution. Hence, the reference set does not come only from top-ranked \llms. For \leetcodeComp, the \llms supply 2,079~reference solutions across all languages and metrics, from 95 (4.6\%) for \deepseek to 250 (12.0\%) for \fable. For \clbgComp, the \llms supply 210~reference solutions, from 5 (2.4\%) for \gptfivemini to 37 (17.6\%) for \gptastra. The appendix gives the full per-\llm, per-language breakdown.

The reference set includes no human-written solutions, because we find no human-written set that fits all our problems. For \leetcode problems, the corpus that we use~\cite{LimYeri2025LeetCodeSolutions} carries human-written solutions only in Python. Additionally, the largest open, multi-language collection~\cite{doocs_leetcode} records no provenance and covers at most 29/99~problems of \leetcodeComp in most of our languages. For \clbgComp, we do not use the human-written programs of \clbg~\cite{clbg} either, because many of them are tuned to the hardware, operating system, and compilers that they were written for (e.g., x86 SIMD intrinsics). Many of them do not build or run unchanged on our host. Given the lack of human-written solutions, a ratio of 1~marks the best solution of the evaluated \llms in both competitions.

\subsection{Statistical Analyses} \label{sec:stat-tests-method}
Our evaluation uses four statistical analyses. We deem a result significant when $p < 0.05$ after Holm's correction~\cite{holm1979} over its family of tests. A family holds the 3~metrics of a competition, the 78~pairs of a metric, or the 4~correlations of one analysis. The appendix describes all tests and results.

\runinhead{Overall Differences} To check whether the \llms differ at all in a metric, we use a permutation test~\cite{Ernst2004Permutation} that we call the \emph{\llm shuffle test}. The test shuffles, inside each cell, which \llm produced which solution, and compares the spread of the Bradley-Terry strengths before and after the shuffle.

\runinhead{Pairwise Differences} To find which pairs of \llms differ, we bootstrap the data~\cite{Efron1979Bootstrap} by drawing 10,000~resamples of the same size with replacement. We then refit the Bradley-Terry model on each, and test the strength gap of each of the 78~pairs of \llms across the resamples. On both competitions, we resample whole problems, because the cells of one problem in different languages are correlated. On \clbgComp, its 10~problems leave these tests little power. \llms that no other \llm beats significantly are \emph{tied leaders}.

\runinhead{Language Effect} To check whether a language changes the order of \llms, we compare the rankings of every two languages with Kendall's~$\tau$~\cite{kendall_tau}. A $\tau$ of $t$ means that two languages order a fraction $(1+t)/2$ of the \llm pairs the same way. A second permutation test, the \emph{language shuffle test}, compares the mean~$\tau$ over the 21~pairs of languages with its value after shuffling the languages of each problem.

\runinhead{Relation to Other Properties} To relate the rankings to other \llm properties (e.g., \passatone and cost), we use Spearman's rank correlation~$\rho$~\cite{spearman}.

\subsection{Evaluation Cost}
To show what it takes to add an \llm to \perfarena, we report the cost of our evaluation campaign, including the generation, validation, and measurement of all solutions. Adding an \llm takes one \emph{sweep}. A sweep submits all problems of a competition in each of its languages. On average, a sweep costs \$8.11 and 22~hours of serial generation and measurement for \leetcodeComp, and \$4.35 and 5~hours for \clbgComp. The cost of a sweep depends strongly on the \llm, from \$0.56 for \gptluna to \$19.56 for \fable on \leetcodeComp, and from \$0.12 for \gptfivemini to \$13.59 for \fable on \clbgComp. Measurement, not generation, takes most of the sweep time. For \leetcodeComp, the validation and measurement of all solutions take 246~hours, 6.1$\times$ the generation time. For \clbgComp, they take 45~hours, 2.3$\times$ the generation time. In total, our campaign spends \$162 on generation (for 19.7~million response tokens), and takes about 15~days to finish. \Cref{tab:cost} gives the full breakdown of the costs per \llm and competition.


\section{Evaluation} \label{sec:results}

To show the analyses that \perfarena enables, we evaluate a range of \llms on both supported competitions. Our evaluation investigates the following research questions:

\begin{itemize}[leftmargin=*]
\item \RQ{1}: Which \llm generates code that has the best \latency, \energy, or \memory?
\item \RQ{2}: Which \llms generate the most performant code when we consider \latency, \energy, and \memory together?
\item \RQ{3}: How does the generation cost of an \llm relate to the performance of its code?
\end{itemize}

\begin{table*}
\centering
\caption{Win probability (\%) on \leetcodeComp and \clbgComp, sorted by \latency win probability on \leetcodeComp. Bold marks tied leaders.}
\label{tab:rq1-overall}
\resizebox{\textwidth}{!}{
\begin{tabular}{l rr rr rr rr}
\toprule
 & \multicolumn{2}{c}{\textbf{Latency}} & \multicolumn{2}{c}{\textbf{Energy}} & \multicolumn{2}{c}{\textbf{Peak memory}} & \multicolumn{2}{c}{\textbf{Latency ratio}} \\
\cmidrule(lr){2-3}\cmidrule(lr){4-5}\cmidrule(lr){6-7}\cmidrule(l){8-9}
\textbf{\llm} & \leetcodeComp & \clbgComp & \leetcodeComp & \clbgComp & \leetcodeComp & \clbgComp & \leetcodeComp & \clbgComp \\
\midrule
\gptastra & \textbf{62} & \textbf{72} & \textbf{61} & \textbf{72} & \textbf{50} & \textbf{65} & 1.73 & 1.52 \\
\opusfive & \textbf{58} & \textbf{63} & \textbf{58} & \textbf{61} & \textbf{51} & 43 & 1.97 & 1.76 \\
\gptsol & 56 & \textbf{59} & 50 & 58 & \textbf{50} & 49 & 1.96 & 1.87 \\
\sonnet & 54 & 44 & 54 & \textbf{52} & \textbf{50} & \textbf{47} & 2.10 & 2.29 \\
\fable & 53 & \textbf{65} & 52 & 63 & \textbf{51} & 48 & 2.00 & 1.76 \\
\glm & 53 & \textbf{61} & 52 & \textbf{58} & \textbf{49} & \textbf{50} & 2.10 & 2.15 \\
\geminiflash & 51 & \textbf{57} & \textbf{56} & \textbf{53} & \textbf{50} & \textbf{50} & 1.94 & 2.14 \\
\gptfivefive & 49 & \textbf{60} & 48 & \textbf{60} & \textbf{49} & 45 & 2.07 & 1.89 \\
\gptfivemini & 47 & \textbf{31} & 47 & \textbf{30} & \textbf{49} & 44 & 1.90 & 5.02 \\
\gptluna & 46 & 25 & 45 & 26 & \textbf{50} & \textbf{51} & 2.10 & 4.38 \\
\deepseek & 43 & 39 & 45 & 38 & \textbf{50} & \textbf{51} & 2.37 & 2.50 \\
\opusfour & 39 & 29 & 43 & 31 & \textbf{50} & \textbf{53} & 2.32 & 3.91 \\
\grok & 39 & 45 & 40 & \textbf{48} & \textbf{50} & \textbf{54} & 2.72 & 2.82 \\
\bottomrule
\end{tabular}}
\end{table*}

\subsection{Correctness of the Generated Code} \label{sec:correctness}
We validate the \leetcodeComp solutions using the \leetcode judge, and the \clbgComp solutions using the custom correctness oracle of \perfarena. The \passatone of an \llm is the share of its cells whose single sample passes the correctness oracle. For \leetcodeComp, \passatone ranges from 88.5\% (\gptfivemini) to 99.7\% (\gptastra). Since 12~\llms solve more than 90\% of the cells, correctness barely separates \llms on \leetcodeComp. For \clbgComp, \passatone ranges from 31.4\% (\glm) to 100\% (\gptastra). Since only 7~\llms solve more than 90\% of the cells, correctness separates \llms on \clbgComp more than on \leetcodeComp.

On \leetcodeComp, an \llm usually solves a problem in every language or in none. Of the 1,287~\llm-problem pairs of \leetcodeComp, 89.7\% receive the same verdict in all languages (\Cref{tab:xlang-consistency} in the appendix), with a per-language rejection rate between 2.6\% (TypeScript) and 4.2\% (C++). On \clbgComp, only 77/130~pairs (59.2\%) receive the same verdict in all languages. The two least correct \llms lose the most on \leetcode Hard problems (\gptfivemini falls from 99\% on Easy to 73\% on Hard, and \gptluna from 99\% to 87\%). The appendix gives the \passatone of each \llm and its full breakdown per language and per difficulty.

Correctness also says little about which \llm generates faster code. On \leetcodeComp, \passatone and the \latency win probability correlate positively, but the link does not survive Holm's correction ($\rho = +0.59$, $p = 0.04$). Additionally, \passatone shows no significant link with \energy ($\rho = +0.40$, $p = 0.18$). On \clbgComp, the \llms with higher \passatone generate code that consumes less energy ($\rho = +0.73$, $p = 0.005$), but their link with \latency does not survive Holm's correction ($\rho = +0.59$, $p = 0.03$).

\begin{figure*}
\centering
\subfloat[\leetcodeComp]{\includegraphics[width=0.495\textwidth]{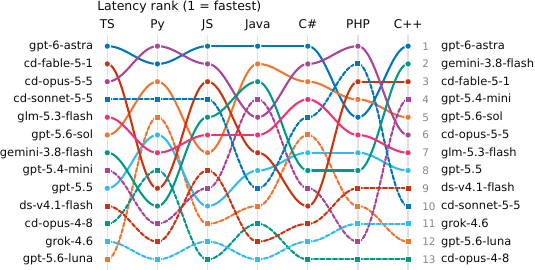}}\hfill
\subfloat[\clbgComp]{\includegraphics[width=0.495\textwidth]{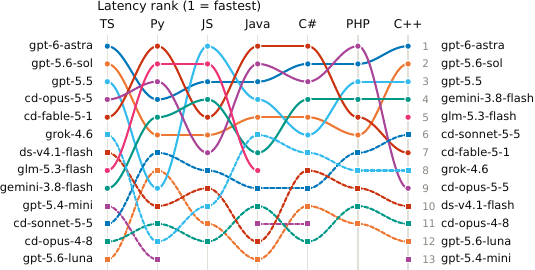}}\\
\subfloat[\leetcodeComp]{\includegraphics[width=0.495\textwidth]{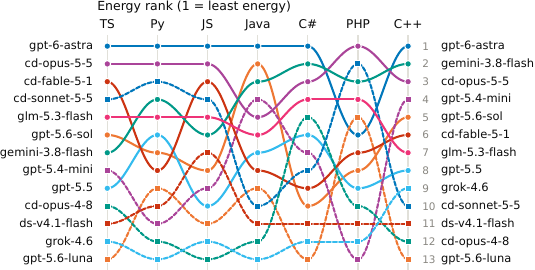}}\hfill
\subfloat[\clbgComp]{\includegraphics[width=0.495\textwidth]{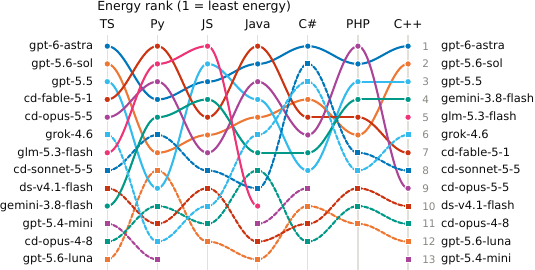}}\\
\subfloat[\leetcodeComp]{\includegraphics[width=0.495\textwidth]{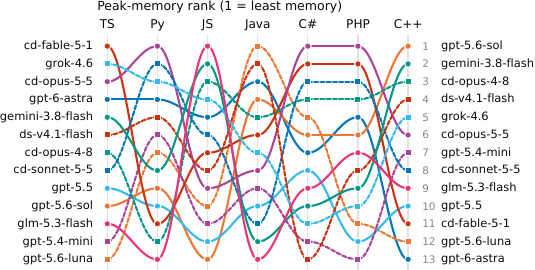}}\hfill
\subfloat[\clbgComp]{\includegraphics[width=0.495\textwidth]{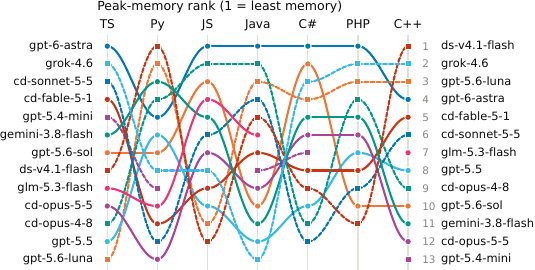}}
\caption{Per-language rank of each \llm by win probability on \latency, \energy, and \memory on \leetcodeComp and \clbgComp. Languages go from the most to least used according to GitHub in 2025~\cite{GitHub2025Octoverse}. The labels abbreviate \code{claude-} as \code{cd-} and \code{deepseek-} as \code{ds-}.}
\label{fig:rank-bump}
\end{figure*}

\subsection{Best \llm per Performance Metric (\RQ{1})} \label{sec:rq-family-size-language}
\Cref{tab:rq1-overall} shows that \gptastra and \opusfive generate the fastest code on \leetcodeComp, with win probabilities of 62\% and 58\%, respectively. No other \llm beats them significantly in terms of \latency. On \clbgComp, \gptastra leads with a 72\% win probability, tied with \fable, \opusfive, \glm, \gptfivefive, \gptsol, \geminiflash, and \gptfivemini. On the cells that an \llm and \gptastra solve, the code of that \llm takes on average $1.14$--$1.61\times$ as long to run on \leetcodeComp and $1.2$--$3.61\times$ on \clbgComp. The slowest code comes from \grok on \leetcodeComp ($1.61\times$), and \gptluna, \opusfour, and \gptfivemini on \clbgComp ($2.6$--$3.6\times$). \gptfivemini is among the tied leaders only because it solves 24/70~cells, too few for the tests to separate it from \gptastra.

\Energy of code closely follows its \latency ($\rho = 0.91$ on \leetcodeComp and $0.97$ on \clbgComp). On \leetcodeComp, \gptastra, \opusfive, and \geminiflash are tied leaders, with win probabilities of 61\%, 58\%, and 56\%, respectively. Of them, only \geminiflash is a tied leader in every language. On \clbgComp, \gptastra leads with a 72\% win probability, tied with \opusfive, \gptfivefive, \glm, \geminiflash, \sonnet, \grok, and \gptfivemini. On \leetcodeComp, the \energy of one call is its \latency times the mean package power of the harness process. This power differs little between solutions. Hence, a solution that runs longer also consumes more energy. However, the fastest solution of a cell also uses the least \energy in 64\% of \leetcodeComp cells and 80\% of \clbgComp cells. On \clbgComp, every \llm \energy ratio also lies below its \latency ratio (13/13~\llms, sign test $p < 0.001$), because the fastest solution often draws more power (e.g., by using both cores).

On \leetcodeComp, \memory does not separate \llms, because each \llm wins 49--51\% of its duels. Neither the pairwise tests nor the \llm shuffle test finds a significant difference ($p = 0.09$). The language runtime may use most of the memory of a test case, leaving the differences between solutions within the noise of a single reading. On \clbgComp, \gptastra uses the least \memory, with a 65\% win probability, tied with 7~\llms.

Although the order of the \llms changes with the language, the tied leaders barely change. On \leetcodeComp, the rankings of two languages ($\tau = 0.50$) agree less than after shuffling the languages of each problem ($\tau = 0.72$, $p < 0.001$). Two \llms lead in at least one language (\opusfive in Python and PHP, and \gptastra in the other languages). However, \gptastra is among the tied leaders in every language, and \opusfive, \sonnet, and \glm in all but one. On \clbgComp, 10~problems per language are too few to separate the \llms within a language. \Cref{fig:rank-bump} shows how the rank of each \llm changes across languages. On \leetcodeComp, the top and the bottom of the \latency ranking change little. For example, \gptastra ranks 1st--5th in every language, \opusfive 1st--6th, and \grok 11th--13th. The \llms in the middle move the most. For example, \geminiflash ranks 2nd in C++ but 10th in Python, and \gptfivemini 4th in C++ but 12th in PHP. On \clbgComp, \gptastra ranks 1st--4th in every language in both \latency and \energy. With respect to \memory, the rank of 10~\llms on \leetcodeComp changes by 9 or more places across languages, as expected when no pair of \llms differs. \Cref{tab:rq1-languages,tab:rq1-languages-energy,tab:rq1-languages-memory} in the appendix give the win probability of every \llm in every language and metric.

The \llm shuffle test confirms that \llms differ in \latency and \energy on both competitions ($p < 0.001$) and in \memory only on \clbgComp ($p = 0.02$). The pairwise tests separate 42/78~pairs in \latency on \leetcodeComp and 21 on \clbgComp. The appendix lists all tests.

\roundbox{On \leetcodeComp, \gptastra and \opusfive tie for the fastest code and share the lead in \energy with \geminiflash. On \clbgComp, \gptastra leads. \Memory barely separates the \llms on either competition.}

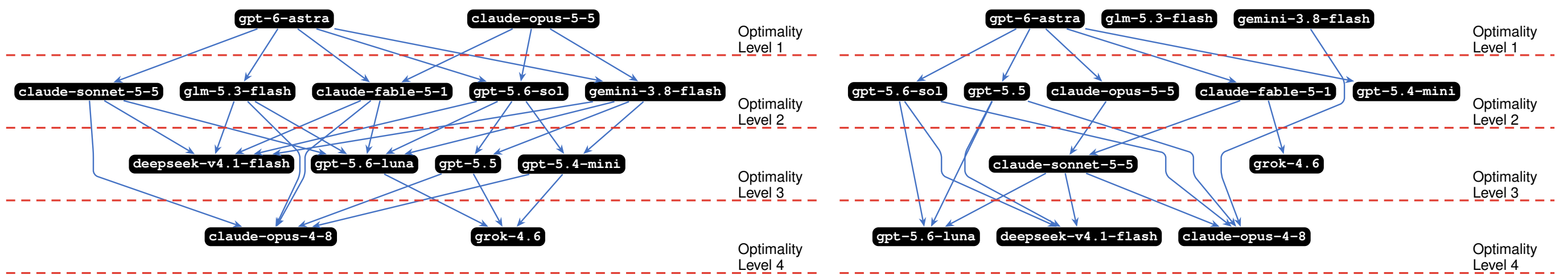
\begin{figure*}
\centering
\subfloat[\leetcodeComp]{
\definecolor{paretoblue}{HTML}{4472C4}\definecolor{paretored}{HTML}{E03C31}
\begin{tikzpicture}[scale=0.9082, transform shape, y=0.7cm,
  pnode/.style={fill=black, text=white, rounded corners=2pt, inner sep=1.5pt, align=center, font=\sffamily\fontsize{5.3}{5.8}\selectfont\bfseries},
  pedge/.style={-{Stealth[length=3pt, width=2.4pt]}, draw=paretoblue, line width=0.45pt, rounded corners=1.5pt},
  plevel/.style={draw=paretored, dash pattern=on 3pt off 2pt, line width=0.6pt},
  plabel/.style={anchor=south west, align=left, inner sep=0.8pt, font=\sffamily\fontsize{5.3}{5.8}\selectfont}]
\node[pnode] (m0) at (3.459,0.000) {\gptastra};
\node[pnode] (m1) at (6.461,-1.327) {\gptsol};
\node[pnode] (m2) at (4.486,-2.654) {\gptluna};
\node[pnode] (m3) at (5.811,-2.654) {\gptfivefive};
\node[pnode] (m4) at (7.136,-2.654) {\gptfivemini};
\node[pnode] (m5) at (4.714,-1.327) {\fable};
\node[pnode] (m6) at (3.287,-3.981) {\opusfour};
\node[pnode] (m7) at (6.647,0.000) {\opusfive};
\node[pnode] (m8) at (0.958,-1.327) {\sonnet};
\node[pnode] (m9) at (2.529,-2.654) {\deepseek};
\node[pnode] (m10) at (8.207,-1.327) {\geminiflash};
\node[pnode] (m11) at (6.322,-3.981) {\grok};
\node[pnode] (m12) at (2.862,-1.327) {\glm};
\draw[pedge] (m0) -- (m1);
\draw[pedge] (m0) -- (m5);
\draw[pedge] (m0) -- (m8);
\draw[pedge] (m0) -- (m10);
\draw[pedge] (m0) -- (m12);
\draw[pedge] (m1) -- (m2);
\draw[pedge] (m1) -- (m3);
\draw[pedge] (m1) -- (m4);
\draw[pedge] (m1) -- (m9);
\draw[pedge] (m2) -- (m11);
\draw[pedge] (m3) -- (m6);
\draw[pedge] (m3) -- (m11);
\draw[pedge] (m4) -- (m6);
\draw[pedge] (m4) -- (m11);
\draw[pedge] (m5) -- (m2);
\draw[pedge] (m5) -- (3.725,-2.454) -- (3.725,-2.854) -- (m6);
\draw[pedge] (m5) -- (m9);
\draw[pedge] (m7) -- (m1);
\draw[pedge] (m7) -- (m5);
\draw[pedge] (m7) -- (m10);
\draw[pedge] (m8) -- (m2);
\draw[pedge] (m8) -- (1.016,-2.454) -- (1.016,-2.854) -- (m6);
\draw[pedge] (m8) -- (m9);
\draw[pedge] (m10) -- (m2);
\draw[pedge] (m10) -- (m3);
\draw[pedge] (m10) -- (m4);
\draw[pedge] (m10) -- (m9);
\draw[pedge] (m12) -- (m2);
\draw[pedge] (m12) -- (3.659,-2.454) -- (3.659,-2.854) -- (m6);
\draw[pedge] (m12) -- (m9);
\draw[plevel] (-0.100,-0.663) -- (10.312,-0.663);
\node[plabel] at (9.232,-0.663) {Optimality\\Level 1};
\draw[plevel] (-0.100,-1.990) -- (10.312,-1.990);
\node[plabel] at (9.232,-1.990) {Optimality\\Level 2};
\draw[plevel] (-0.100,-3.317) -- (10.312,-3.317);
\node[plabel] at (9.232,-3.317) {Optimality\\Level 3};
\draw[plevel] (-0.100,-4.644) -- (10.312,-4.644);
\node[plabel] at (9.232,-4.644) {Optimality\\Level 4};
\end{tikzpicture}}\hfill
\subfloat[\clbgComp]{
\definecolor{paretoblue}{HTML}{4472C4}\definecolor{paretored}{HTML}{E03C31}
\begin{tikzpicture}[scale=0.9082, transform shape, y=0.7cm,
  pnode/.style={fill=black, text=white, rounded corners=2pt, inner sep=1.5pt, align=center, font=\sffamily\fontsize{5.3}{5.8}\selectfont\bfseries},
  pedge/.style={-{Stealth[length=3pt, width=2.4pt]}, draw=paretoblue, line width=0.45pt, rounded corners=1.5pt},
  plevel/.style={draw=paretored, dash pattern=on 3pt off 2pt, line width=0.6pt},
  plabel/.style={anchor=south west, align=left, inner sep=0.8pt, font=\sffamily\fontsize{5.3}{5.8}\selectfont}]
\node[pnode] (m0) at (2.401,0.000) {\gptastra};
\node[pnode] (m1) at (0.641,-1.327) {\gptsol};
\node[pnode] (m2) at (1.007,-3.981) {\gptluna};
\node[pnode] (m3) at (1.913,-1.327) {\gptfivefive};
\node[pnode] (m4) at (7.153,-1.327) {\gptfivemini};
\node[pnode] (m5) at (5.353,-1.327) {\fable};
\node[pnode] (m6) at (5.080,-3.981) {\opusfour};
\node[pnode] (m7) at (3.396,-1.327) {\opusfive};
\node[pnode] (m8) at (2.761,-2.654) {\sonnet};
\node[pnode] (m9) at (2.964,-3.981) {\deepseek};
\node[pnode] (m10) at (5.841,0.000) {\geminiflash};
\node[pnode] (m11) at (5.619,-2.654) {\grok};
\node[pnode] (m12) at (3.990,0.000) {\glm};
\draw[pedge] (m0) -- (m1);
\draw[pedge] (m0) -- (m3);
\draw[pedge] (m0) -- (m4);
\draw[pedge] (m0) -- (m5);
\draw[pedge] (m0) -- (m7);
\draw[pedge] (m1) -- (m2);
\draw[pedge] (m1) -- (4.073,-2.454) -- (4.073,-2.854) -- (m6);
\draw[pedge] (m1) -- (1.203,-2.454) -- (1.203,-2.854) -- (m9);
\draw[pedge] (m3) -- (m2);
\draw[pedge] (m3) -- (4.427,-2.454) -- (4.427,-2.854) -- (m6);
\draw[pedge] (m3) -- (1.503,-2.454) -- (1.503,-2.854) -- (m9);
\draw[pedge] (m5) -- (m8);
\draw[pedge] (m5) -- (m11);
\draw[pedge] (m7) -- (m8);
\draw[pedge] (m8) -- (m2);
\draw[pedge] (m8) -- (m6);
\draw[pedge] (m8) -- (m9);
\draw[pedge] (m10) -- (6.359,-1.127) -- (6.359,-1.527) -- (4.782,-2.454) -- (4.782,-2.854) -- (m6);
\draw[plevel] (-0.100,-0.663) -- (9.047,-0.663);
\node[plabel] at (7.967,-0.663) {Optimality\\Level 1};
\draw[plevel] (-0.100,-1.990) -- (9.047,-1.990);
\node[plabel] at (7.967,-1.990) {Optimality\\Level 2};
\draw[plevel] (-0.100,-3.317) -- (9.047,-3.317);
\node[plabel] at (7.967,-3.317) {Optimality\\Level 3};
\draw[plevel] (-0.100,-4.644) -- (9.047,-4.644);
\node[plabel] at (7.967,-4.644) {Optimality\\Level 4};
\end{tikzpicture}}
\caption{Pareto optimality levels of the \llms across all performance metrics. An arrow points from an \llm to each \llm that it directly dominates.}
\label{fig:pareto-levels}
\end{figure*}

\subsection{Best \llms Across All Performance Metrics (\RQ{2})} \label{sec:rq-pareto}
Since our performance metrics lack a common unit and their priorities vary by deployment (e.g., low \latency for cloud services, low \energy for mobile devices, or low \memory for containers), a single composite score would unfairly favor specific use cases. Therefore, to compare the \llms neutrally, we use Pareto dominance~\cite{ehrgott_multicriteria}. An \llm dominates another if it is significantly better in at least one metric and significantly worse in none. A difference that the pairwise tests do not find significant counts as a tie. Hence, measurement noise cannot decide dominance. A dominated \llm is inferior under any weighting of the metrics. Hence, the non-dominated \llms include the best \llm for any deployment, without fixing weights in advance.

\Cref{fig:pareto-levels} sorts \llms into optimality levels. Level~1 holds the non-dominated \llms. Level~2 holds the \llms that only Level~1 dominates, and arrows mark direct dominance. \gptastra dominates 11~other \llms on \leetcodeComp and 10~on \clbgComp. \rr{On \clbgComp, only 5 of the 78~pairs of \llms differ significantly in \memory, and \gptastra uses less \memory in all 5.} \rr{\gptastra dominates \opusfive, \gptfivefive, and \gptfivemini through these \memory differences alone.} On \leetcodeComp, \opusfive shares Level~1 with \gptastra, because no metric separates them significantly. \sonnet falls to Level~2, because \gptastra beats it significantly in \latency and \energy. On \clbgComp, \glm and \geminiflash share Level~1 with \gptastra, because no metric separates them significantly. \glm solves only 22/70~cells, too few for the tests to separate it from \gptastra. \Cref{fig:pareto-levels} shows that \gptastra is the only \llm on Level~1 of both competitions, because no other \llm significantly beats it on any metric. \rr{Without \memory, \opusfive would also reach Level~1 on \clbgComp, and so on both competitions.} The lowest level holds \grok and \opusfour on \leetcodeComp, and \opusfour, \gptluna, and \deepseek on \clbgComp.

\roundbox{On both \leetcodeComp and \clbgComp, \gptastra is the only \llm that no other \llm dominates.}

\begin{table*}
\centering
\caption{Correctness, generation cost, and time of the sweep of each \llm on \leetcodeComp (LC) and \clbgComp (CLBG), sorted by USD per correct solution on LC. The last row gives the overall rates and sums. Bold marks the best value in each column.}
\label{tab:cost}
\resizebox{\textwidth}{!}{
\begin{tabular}{l rr rr rr rr rr rr rr}
\toprule
 & \multicolumn{2}{c}{\textbf{\passatone (\%)}} & \multicolumn{4}{c}{\textbf{Response tokens}} & \multicolumn{4}{c}{\textbf{Generation Cost (USD)}} & \multicolumn{4}{c}{\textbf{Time (h)}} \\
\cmidrule(lr){2-3}\cmidrule(lr){4-7}\cmidrule(lr){8-11}\cmidrule(lr){12-15}
 & \multicolumn{2}{c}{} & \multicolumn{2}{c}{\textbf{Per Solution}} & \multicolumn{2}{c}{\textbf{Sweep}} & \multicolumn{2}{c}{\textbf{Per Solution}} & \multicolumn{2}{c}{\textbf{Sweep}} & \multicolumn{2}{c}{\textbf{Generation}} & \multicolumn{2}{c}{\textbf{Measurement}} \\
\cmidrule(lr){4-5}\cmidrule(lr){6-7}\cmidrule(lr){8-9}\cmidrule(lr){10-11}\cmidrule(lr){12-13}\cmidrule(lr){14-15}
\textbf{\llm} & \textbf{LC} & \textbf{CLBG} & \textbf{LC} & \textbf{CLBG} & \textbf{LC} & \textbf{CLBG} & \textbf{LC} & \textbf{CLBG} & \textbf{LC} & \textbf{CLBG} & \textbf{LC} & \textbf{CLBG} & \textbf{LC} & \textbf{CLBG} \\
\midrule
\code{gpt-5.6-luna} & 93.1 & 74.3 & 626 & 3,837 & 404K & 200K & \textbf{0.0009} & \textbf{0.0047} & \textbf{0.56} & 0.25 & 1.6 & 0.7 & 21.3 & 5.8 \\
\code{gpt-5.4-mini} & 88.5 & 34.3 & \textbf{307} & 1,942 & \textbf{188K} & \textbf{47K} & \textbf{0.0009} & 0.0049 & 0.57 & \textbf{0.12} & \textbf{0.5} & \textbf{0.1} & 18.4 & 2.5 \\
\code{deepseek-v4.1-flash} & 97.4 & 64.3 & 2,323 & 42,035 & 1,568K & 1,892K & 0.0033 & 0.0559 & 2.25 & 2.51 & 3.1 & 2.2 & 19.0 & 9.2 \\
\code{gemini-3.8-flash} & 95.8 & 92.9 & 1,190 & 13,235 & 790K & 860K & 0.0049 & 0.0501 & 3.28 & 3.25 & 1.9 & 1.5 & 18.6 & 2.3 \\
\code{claude-sonnet-5-5} & 98.1 & 97.1 & 586 & 1,760 & 398K & 120K & 0.0075 & 0.0193 & 5.11 & 1.31 & 0.9 & 0.3 & 22.1 & 1.2 \\
\code{gpt-6-astra} & \textbf{99.7} & \textbf{100.0} & 321 & 1,458 & 222K & 102K & 0.0109 & 0.0391 & 7.55 & 2.74 & 1.5 & 0.7 & 19.9 & 1.8 \\
\code{gpt-5.6-sol} & 98.7 & 98.6 & 474 & 3,020 & 324K & 208K & 0.0118 & 0.0626 & 8.09 & 4.32 & 1.5 & 0.9 & 18.7 & 1.1 \\
\code{claude-opus-4-8} & 96.1 & 78.6 & 368 & \textbf{1,306} & 245K & 72K & 0.0134 & 0.0378 & 8.93 & 2.08 & 0.9 & 0.2 & 23.1 & 2.5 \\
\code{glm-5.3-flash} & 97.1 & 31.4 & 6,494 & 161,627 & 4,371K & 3,556K & 0.0148 & 0.3642 & 9.98 & 8.01 & 15.0 & 9.0 & \textbf{9.6} & 6.0 \\
\code{claude-opus-5-5} & 99.0 & 97.1 & 584 & 2,286 & 401K & 155K & 0.0150 & 0.0491 & 10.27 & 3.34 & 1.3 & 0.4 & 20.5 & \textbf{0.9} \\
\code{grok-4.6} & 98.4 & 90.0 & 2,332 & 6,974 & 1,590K & 439K & 0.0156 & 0.0435 & 10.62 & 2.74 & 8.0 & 1.7 & 14.7 & 7.6 \\
\code{gpt-5.5} & 98.4 & 94.3 & 817 & 6,100 & 557K & 403K & 0.0274 & 0.1858 & 18.71 & 12.27 & 2.5 & 1.5 & 18.4 & 1.2 \\
\code{claude-fable-5-1} & 98.0 & 97.1 & 410 & 3,839 & 278K & 261K & 0.0288 & 0.1999 & 19.56 & 13.59 & 1.3 & 0.9 & 21.2 & 3.4 \\
\midrule
\textbf{All} & 96.8 & 80.8 & 1,300 & 11,312 & 11,337K & 8,314K & 0.0121 & 0.0769 & 105.49 & 56.53 & 40.1 & 20.1 & 245.6 & 45.4 \\
\bottomrule
\end{tabular}}
\end{table*}

\subsection{\llm Cost and Performance (\RQ{3})} \label{sec:rq-tradeoffs}
We measure the generation cost of an \llm in two units: response tokens and USD. Both units include the reasoning tokens, which providers bill as output tokens. To relate cost to performance independently of correctness, we divide the cost of a sweep by the number of attempted solutions. To reflect what a developer pays, we also report the cost per correct solution. This cost grows when an \llm fails often. 

\Cref{tab:cost} lists the cost per correct solution and per sweep of each \llm. The table shows that \gptastra emits few response tokens per correct solution on both competitions (321 on \leetcodeComp and 1,458 on \clbgComp). Only \gptfivemini on \leetcodeComp (307) and \opusfour on \clbgComp (1,306) emit fewer. However, its price per token is among the highest. A correct \gptastra solution costs \$0.011 on \leetcodeComp and \$0.039 on \clbgComp. Both \gptluna and \gptfivemini cost 8--12$\times$ less per correct solution, but their code takes $1.23\times$ as long to run as that of \gptastra on \leetcodeComp, and $2.76\times$ and $3.61\times$, respectively, on \clbgComp. Conversely, per correct solution, the three most expensive \llms cost 2.6$\times$ (\fable), 2.5$\times$ (\gptfivefive), and 1.4$\times$ (\grok) as much as \gptastra on \leetcodeComp and 9.3$\times$ (\glm), 5.1$\times$ (\fable), and 4.7$\times$ (\gptfivefive) on \clbgComp. Their code still runs 1.19--1.61$\times$ as long on \leetcodeComp and 1.21--1.29$\times$ on \clbgComp.

The cost of an \llm relates to the performance of its code only on \clbgComp. There, the \llms that cost more per attempted solution in USD generate faster code ($\rho = +0.81$ with the \latency win probability, $p < 0.001$). The four cheapest \llms (\gptfivemini, \gptluna, \sonnet, and \opusfour) drive this link (without them, $\rho = +0.53$, $p = 0.14$), because they include the slowest three (\gptluna, \opusfour, and \gptfivemini). We find no significant relation between the token count and \latency on either competition, or between either cost and \latency on \leetcodeComp ($|\rho| \le 0.33$, $p \ge 0.27$). The appendix relates the cost to the language of the solution.

\roundbox{Paying more does not buy faster code. Only on \clbgComp do costlier \llms generate faster code, and only because the cheapest \llms are also the slowest. \gptastra costs less than half as much per correct solution as the two most expensive \llms on each competition, yet their code is not faster.}

\section{Threats to Validity}
\label{sec:threats}

\subsection{Construct Validity}
\label{sec:threats-construct}

\runinhead{C1} \Latency and \energy produce similar \llm rankings ($\rho = 0.91$ on \leetcodeComp and $0.97$ on \clbgComp), because the per-invocation \energy of a \leetcodeComp solution is its per-invocation \latency times the average package power of its harness process, which varies little (13.1--16~W for 90\% of the solutions). However, we report the three metrics separately, because they still disagree on specific tasks and languages.

\runinhead{C2} We rank \leetcode problems by the ratio of their 95\textsuperscript{th} to their 10\textsuperscript{th} \latency percentile, which a 10\textsuperscript{th} percentile near the timing floor of the \leetcode judge may inflate. Our absolute criteria (a median and a percentile gap of at least 10~ms each) limit this bias. Moreover, these distributions come from accepted Python~3 submissions only. Hence, a problem may separate fast from slow solutions less in another language. In the median cell, the slowest correct solution takes 2.12$\times$ as long as the fastest in Python, but 1.75$\times$ in C\#, the language with the smallest spread. Every language still separates the solutions, and the problems with the largest spread in Python tend to have a large spread in the other languages too ($\rho = 0.53$--$0.73$).

\runinhead{C3} An oracle that only compares outputs cannot tell whether a solution does the intended work in its own language. Such shortcuts win every duel in their cell. Hence, \perfarena flags for manual review any accepted solution that runs at least 2$\times$ faster than the median of the other solutions in its cell, although this flag may miss a moderately faster shortcut. On \clbgComp, the flag finds 5~shortcuts. One Python \bench{n-body} solution of \opusfive compiles C code and calls it through a foreign-function interface, running 64$\times$ faster than the next fastest solution. Four \bench{binary-trees} solutions (\deepseek in PHP, Python, and TypeScript, and \gptfivefive in Python) compute the checksums by formula, never building a tree, running 22--262$\times$ faster than the fastest solution that does. We exclude all 5~from the rankings and the reference set, but count them for \passatone.

\runinhead{C4} Each language has its own libraries and harness (JavaScript and TypeScript share one Node.js measuring loop). For example, the arbitrary-precision arithmetic available to \bench{pidigits} differs between languages, and only some harnesses copy the input before every call. Our rankings compare \llms only within one language, where all \llms share the same libraries and harness. However, absolute values across languages may partly reflect these differences rather than the generated code.

\runinhead{C5} The reference set holds only \llm-generated solutions. Hence, a ratio of $1.00$ marks the best solution of the evaluated \llms, not the best known solution.

\runinhead{C6} On \leetcodeComp, \perfarena averages over a few kept test cases instead of all test cases of a problem. On the profiling data, the two averages differ by a median of 4.1\% (mean 9.2\%) and by up to 67\% (\bench{random-pick-index}). However, only the order of solutions within a cell affects our rankings. We measure 10~random problems that all \llms solve in both Python and C++ on all their profiled test cases (53--100 per problem, once for 0.3~s each). The single short run adds noise to both averages, making the check conservative. Over the 1,560~duels of these 20~cells, 85\% in \latency and 83\% in \energy keep their outcome. Most others change between a tie and a win. Within a cell, the two orders of the \llms agree at a median $\tau$ of 0.79. Hence, a ranking from all test cases could differ from ours in close duels, but would rarely reverse one.

\subsection{Internal Validity}
\label{sec:threats-internal}

\runinhead{I1} Other processes, frequency scaling, thermal throttling, and drift of the host add noise to every reading. Hence, the runner measures one solution at a time, refuses to start while any core is more than 60\% busy, and repeats each \latency and \energy measurement seven times. The \energy readings also include the idle power of the package, a median of 0.43~W over 10~minutes with nothing running. This idle power makes up 2.7--3.3\% of the \energy of 90\% of the \leetcodeComp solutions and 1.8--2.9\% on \clbgComp, changing \energy ratios by less than the tie margins.

\runinhead{I2} Process startup may dominate a short run, and its cost differs between languages. However, every solution in a language pays about the same cost, which affects absolute values more than the ranking of the \llms.

\runinhead{I3} Repeated calls on one input measure a warm state that a single execution never reaches, and may let a solution or a compiler skip work. The Java, C\#, and Python harnesses reuse one solution object across calls. Hence, a solution that caches its results may look much faster (e.g., 31~ns per call in Java against 2.5~$\mu$s in C++ on one problem). The checksums prevent \dce but not caching, and a \jit compiler may still move a call out of the loop. A new object for each call, as in the C++, JavaScript, TypeScript, and PHP harnesses, removes most caching but adds object-creation cost, and static memos still survive. A scan of all 8,720 accepted \leetcodeComp solutions, with every hit reviewed by hand, finds 31~solutions in 3~problems that keep a memo between calls (16~in Java, 13~in C\#, 1~in Python, and 1~in PHP). The 2$\times$ review flag catches only 13~of them, because most Java and C\# solutions of one problem memoize, keeping the median of their cells fast. Without these 31~solutions, the \latency and \energy rankings agree with ours at $\tau$ of 0.97 and 1.00. Re-running the 30 Java, C\#, and Python ones with a new object per call slows them by 30$\times$ (median) and the rest of their cells by 1.01$\times$, and the rankings still agree at $\tau = 0.97$ in both metrics. Neither check changes the tied leaders or the optimality levels.

\runinhead{I4} A correctness oracle may itself be wrong. Our custom oracle validates each \clbgComp program at small arguments, while the measured runs at the calibrated argument check only the exit status. Hence, an error that appears only at the larger size (e.g., integer overflow) may go undetected. We run each of the 735 accepted \clbgComp solutions once at its calibrated argument and compare its output with that of the C++ reference implementation under the rules of the oracle, and 733 match. The other two, the \bench{n-body} solutions of \grok in Java and PHP, start Jupiter at a wrong position, which changes the initial energy by $7 \times 10^{-8}$ (below the tolerance of $10^{-6}$) but the final energy by $1.4 \times 10^{-6}$ after 15~million steps. We exclude them like the shortcuts above, and they still count as correct for \passatone.

\runinhead{I5} Eleven of the 13~\llms sample at a nonzero temperature. Hence, a rerun may change some of their solutions and duels. Each ranking rests on up to 99~problems per language, which averages out part of this variation, but we do not resample to measure it.

\runinhead{I6} A fixed warmup may end before a \jit compiler finishes optimizing a solution~\cite{Barrett2017Warmup}. Tripling the budget to 4.5~s changes the median \latency of one call of \gptastra by at most 1.3\% in every language except C\#, where it falls by a median of 3.1\% (up to 9.1\%). For the C\# solutions of all \llms on the same 10~problems, measured 3~times at each budget, the rankings agree at $\tau$ of 0.9 in \latency and 0.97 in \energy. No \llm moves by more than one rank in \latency or two in \energy. Single-run rankings agree about as much across budgets as within one (median $\tau$ of 0.82 and 0.83). The \energy of one call falls by a median of 1.9\% (4.7\% in C\#), mostly because a longer budget dilutes the runtime startup in the average power, a shift that every solution of a language shares.

\runinhead{I7} On \clbgComp, the pairwise tests resample only 10~problems and have little power. Hence, a tied leader there is an \llm that the tests cannot separate from \gptastra, not necessarily one as efficient (e.g., \gptfivemini ties for the lead in \latency with a 31\% win probability on only 24/70~cells). Resampling cells instead separates more pairs (33 instead of 21 of the 78~pairs in \latency, 23 instead of 17 in \energy, and 6 instead of 5 in \memory), but ignores that the cells of one problem are correlated. Under both resamplings, \gptastra stays the only \llm among the tied leaders in every metric on both competitions.

\subsection{External Validity} \label{sec:threats-external}

\runinhead{E1} Every measurement comes from one machine and one energy tool. Hence, a ranking may generalize only to comparable hardware. The relative efficiency of solutions can change across hardware platforms~\cite{Apsan2026EnergyEfficientCode,Solovyeva2025Forge} (e.g., when one solution benefits more than another from more cores, a larger cache, or wider vector units). A different host affects all \llms, but not necessarily by the same amount.

\runinhead{E2} Our languages, problems, and \llms are a snapshot of an arena that keeps growing. \perfarena stores a dated snapshot of each leaderboard. Hence, a later reader may compare a published ranking with its state on that date.

\runinhead{E3} The \leetcode problems and the \clbg programs are public, together with many of their fastest solutions. Hence, they are almost certainly in the training data of the evaluated \llms, which may reproduce a memorized efficient solution rather than derive it. Although all \llms face the same exposure, the results may overstate the efficiency of the code that an \llm generates for new problems.

\section{Related Work}
\label{sec:related-work}

\subsection{Benchmarks and Arenas}
A benchmark fixes a task set and a scoring protocol, enabling comparable results across \llms, as MMLU and GSM8K do for knowledge and reasoning~\cite{Hendrycks2021MMLU,OpenAI2021GSM8K}. For code generation, HumanEval~\cite{Chen2021HumanEval} established a function-level, unit-tested setting and a binary pass/fail notion of correctness that EvalPlus~\cite{Liu2023EvalPlus} and LiveCodeBench~\cite{Jain2025LiveCodeBench} retain. The former strengthened the test suites, while the latter refreshed the task set with newly released problems. NoFunEval~\cite{nofuneval_microsoft} broadens the requirements under test beyond functional correctness to efficiency, security, and maintainability, and finds that \llms with strong functional results generally falter on these non-functional requirements.

Arenas extend a benchmark with continuous updating and relative ranking. Chatbot Arena~\cite{Chiang2024ChatbotArena} collects crowdsourced pairwise votes and converts them into a ranking, using human preference as a judge. MathArena~\cite{Balunovic2025MathArena} instead evaluates \llms automatically against task-specific ground truth and continually incorporates newly released competitions, using freshness as a defense against data contamination. Copilot Arena~\cite{Chi2025CopilotArena} ranks \llms by which code completions developers accept, without executing the code or measuring its performance.

\perfarena keeps the fixed task set and protocol, but replaces the pass/fail verdict with a measurement of the \latency, \memory, and \energy of each accepted solution. Similar to MathArena, \perfarena scores submissions automatically and updates its ranking as new \llms are released. Unlike MathArena, it draws on long-standing public problems whose efficient solutions are also public. Hence, an \llm may reproduce a memorized solution. Similar to Chatbot Arena, our evaluation ranks \llms by head-to-head wins with a Bradley-Terry model. A measurement of two solutions of the same problem in the same language decides each duel instead of a human vote.

\subsection{Efficiency of \llm-Generated Code} \label{sec:related-generated-efficiency}
Empirical studies establish that \llm-generated programs vary widely in efficiency. \citet{Coignion2024Performance} compare the \latency of \llm-generated solutions. \citet{Islam2025Energy}, \citet{tom_ana_llms_python}, and \citet{Apsan2026EnergyEfficientCode} measure the \energy of \llm-generated solutions, showing that they differ substantially from human-written code in \energy and that prompting does not consistently reduce \energy. \citet{Apsan2026EnergyEfficientCode} further show that the relative efficiency of \llm-generated and human-written solutions may change across hardware platforms. Similarly, \citet{Solovyeva2025Forge} find substantial variation across \llms, programming languages, and platforms, with generated solutions sometimes matching or outperforming efficient human-written solutions and performing substantially worse in other settings. \citet{Cursaru2024CodeLlama} measure the \energy of Code Llama solutions with a hardware power meter and find that asking for energy-efficient code gives equal or worse efficiency. \citet{Vartziotis2024GreenCode} compare the \latency, \memory, and \rapl counters for code from 3~AI coding assistants with human solutions, and find the generated code slightly less energy-efficient than optimized human code. \citet{Niu2024Efficiency} measure the \latency of generated Python and C++ code, and find that correctness and efficiency are not positively related.

Several benchmarks standardize this efficiency evaluation. EffiBench~\cite{EffiBench} measures \latency and \memory, while Mercury~\cite{Mercury} and ENAMEL~\cite{ENAMEL} evaluate \latency efficiency. ENAMEL further introduces the efficiency-aware eff@$k$ metric and expert-level reference solutions. EvalPerf~\cite{EvalPerf} scores a solution using executed-instruction counts relative to reference solutions at distinct efficiency levels, avoiding much of the instability of \latency at the cost of a proxy that does not directly measure \latency, \memory, or \energy. COFFE~\cite{Peng2025COFFE} also counts CPU instructions, on stress test cases that an \llm generates, while PIE~\cite{Shypula2024PIE} times C++ programs in the gem5 simulator, because the authors measured a false mean speedup of $1.12\times$ between identical programs on real hardware. ECCO~\cite{Waghjale2024ECCO} measures the \latency and \memory of Python code edits in a fixed cloud image. Instruction counts and simulation give stable values, but they report neither the \energy nor the \memory of a real machine. Therefore, \perfarena measures on real hardware and controls the noise through a quiet host, warmup, and repeated runs. Most of these function-level benchmarks evaluate a single language, typically Python. EffiBench-X~\cite{Qing2025EffiBenchX} extends this evaluation to 6~languages. EnergyBench~\cite{Ionescu2026EnergyBench} jointly considers functional correctness and energy efficiency, but organizes its analysis around prompt-induced variation rather than a persistent cross-model efficiency ranking. GSO~\cite{gso_challenging_optimization}, SWE-Perf~\cite{SWEPerf}, and SWE-fficiency~\cite{SWEfficiency} take the evaluation to real-world repositories and measure whether \llms produce correctness-preserving, \latency-improving patches for existing projects. GSO spans several languages and reports that \llms particularly struggle with low-level languages (e.g., C/C++). KernelBench~\cite{Ouyang2025KernelBench} asks \llms to generate faster GPU kernels and times them against PyTorch. Beyond evaluation, \citet{Afterburner} show that reinforcement learning with execution feedback may improve the \latency and \memory efficiency of an \llm-generated solution, but only for Python and without measuring \energy. EffiLearner~\cite{Huang2024EffiLearner} feeds \latency and \memory profiles back to the \llm to optimize Python code, and Supersonic~\cite{Chen2024Supersonic} trains a small model to generate C/C++ optimizations that the Codeforces judge times.

Our finding that equally correct \llms differ in efficiency agrees with EvalPerf~\cite{EvalPerf} and COFFE~\cite{Peng2025COFFE}, which find that correct code is often inefficient and that larger \llms are not clearly more efficient. \perfarena adds \energy and multiple languages to this picture, showing that the fastest \llm changes with the language even where correctness carries across languages. Similar to \citet{Niu2024Efficiency}, we find no significant link between correctness and \latency, although on \clbgComp, the \llms with higher \passatone generate code that consumes less energy.

Unlike prior efficiency benchmarks, \perfarena measures \energy alongside \latency and \memory for every accepted solution. It compares solutions only within a cell, recording the host, energy source, and measurement settings (including compiler flags) behind every measurement. Its curated workload also clearly separates solutions. In the median cell, the slowest correct solution takes $2.1\times$ as long as the fastest on \leetcodeComp and $4.3\times$ on \clbgComp, while the 7~runs of one solution differ by a median of only 1.3\% and 2.8\% (coefficient of variation).

\subsection{Cross-Language Energy Measurement and Workload Selection}
Implementations of the same programming problems in different languages may exhibit substantial performance differences. \citet{Pereira2017Energy,scp21} measure the \latency, \memory, and \energy of \clbg across several languages using \rapl counters~\cite{rapl}. The resulting ranking separates the fastest/slowest languages by almost two orders of magnitude in \energy. For an arena that evaluates \llms across several languages, this effect may dominate a raw aggregate. This is because an \llm may rank highly simply by succeeding more often in a language whose implementations are relatively efficient. Hence, our evaluation compares \llms only within a cell before aggregating its results.

\citet{easy_green_ase25} revisit these measurements with a causal model that separates a language from its implementation, controlling for the specific program implementation, the number of active cores, and memory activity. Under these controls, the authors find that the choice of language implementation has no significant effect on \energy beyond \latency. \perfarena does not test this claim. Instead, it reports \latency, \memory, and \energy separately for every solution, letting observers see the tasks and languages on which the metrics agree and those on which they diverge.

To keep a full measurement campaign tractable, \perfarena reduces each problem to a representative set of measurement cases with a fingerprint-and-cluster procedure adapted from SimPoint~\cite{Sherwood2002SimPoint}. SimPoint characterizes execution intervals by program behavior, clusters similar intervals, and selects representative points from those clusters. \perfarena extends this idea to cross-language benchmarking by fingerprinting on a single anchor language (i.e., Python), then reuses the same reduced inputs in every other language. The same inputs give every language the same problem instance, although the work that a solution does on them still depends on its algorithm and language harness.

\section{Extending and Reproducing \perfarena}
\label{sec:reproducibility}

\perfarena is an open-source infrastructure with a public instance that lets the community consult its results. Others may deploy their own instance, or extend \perfarena with new competitions or languages. The platform is not tied to one problem set or to one way of producing code. The correctness oracle, the performance oracle and its profiler, the prompt template, and the code-generation workflow are all pluggable. Therefore, a new judging backend, a different energy tool, or an agent workflow plugs in without changing the rest of the platform.

Our reference set also lowers the cost of comparing against our results. For each problem and language, it records, separately for \latency, \energy, and \memory, the best value that any evaluated \llm reaches and the \llm that reaches it. We also version these records. A future study on the same host and energy tool requires only one comparison against this reference set, which costs one sweep per \llm, not a full rerun of the campaign. A study on another host must re-measure all released solutions to rank a new \llm, and only the reference solutions to compute ratios, because measurements from different machines or energy tools are not comparable. To place a new \llm in the ranking, its solutions also duel with the released solutions of the evaluated \llms, and a refit of the Bradley-Terry model ranks it, similar to how a new \llm enters Chatbot Arena~\cite{Chiang2024ChatbotArena}. The win probabilities of the other \llms may then shift slightly.

We release all code, prompts, generated solutions, per-problem measurements, and rankings. The source code of \perfarena is at \url{https://github.com/perfarena/perfarena}, together with guides to deploy an instance and run a campaign. The \perfarena leaderboard is live at \url{https://perfarena.net}. The code of \perfarena is under the Apache License~2.0. The measurement framework of \citet{Pereira2017Energy} that it extends is under the MIT license. The data that \perfarena produces (i.e., the generated solutions, the measurements, the rankings, and the prompt templates) is under the Creative Commons Attribution~4.0 license. These licenses do not cover third-party content. The \leetcode problem statements, which the released prompts contain, remain the property of \leetcode. The \leetcodeComp test cases come from \leetcodeDataset~\cite{Xia2025LeetCodeDatasetHF}, which is under the Apache License~2.0. The \clbgComp reference implementations come from \clbg~\cite{clbg}, which is under the Revised BSD license. Anyone may consult or download the datasets that \perfarena produces to check a leaderboard, to train or fine-tune \llms for efficiency, or to build a new study on top of it. Hence, no result in this paper depends on access to our servers.

\section{Conclusion}
\label{sec:conclusion}

\perfarena is a continuously updated arena that ranks \llms by the \latency, \memory, and \energy of their correct code, under a shared protocol that holds the problem and the language constant in every comparison. Across \leetcodeComp and \clbgComp, \gptastra is the only \llm that no other \llm beats significantly in any metric. Our results show that correct code is not necessarily efficient code. Although 12/13~\llms solve more than 90\% of the \leetcodeComp cells (i.e., a problem in one language), the correct solutions of the slowest \llm take on average 1.6$\times$ as long to run as those of \gptastra. On the heavier \clbgComp programs, this gap grows to 3.6$\times$. Moreover, we find that on \leetcodeComp, an \llm usually solves a problem in every language or in none, but the fastest \llm changes with the language. On \clbgComp, only 59\% of the \llm-problem pairs receive the same verdict in all languages. In practice, developers should compare \llms per language and per metric instead of relying on pass rates or prices. \perfarena supplies the measurements to achieve that goal.

\onecolumn
\appendices
\crefalias{section}{appendix}
\section{Per-Language Efficiency Rankings}
\label{sec:appendix-rankings}

The paper ranks the \llms over all languages of each competition. Here we further report, for each language, the \llms ranked by their win probability in \latency, \energy, and \memory, from a Bradley--Terry model that we fit to the duels of that language only. An \llm with fewer than three solved problems in a language gets no rank there. On \clbgComp, each language holds only ten problems, so most \llms tie for the lead.

\Cref{tab:rq1-languages,tab:rq1-languages-energy,tab:rq1-languages-memory} give the win probability of each \llm in each language on both competitions, in \latency, \energy, and \memory. The tables that follow rank the \llms in each language of both competitions. In each of these ranking tables, subtable~(a) gives \leetcodeComp and subtable~(b) gives \clbgComp. Bold marks the tied leaders, and a dash marks an \llm with no rank in that language.

\begin{table}[H]
\centering
\footnotesize
\caption{\Latency win probability (\%) on \leetcodeComp (LC) and \clbgComp (CLBG), sorted in a descending order of \latency win probability on LC. The highest win probability in each column is in bold. A dash marks an \llm with fewer than 3~solved problems in the corresponding language.}
\label{tab:rq1-languages}
\begin{tabular}{l rr rr rr rr rr rr rr}
\toprule
 & \multicolumn{2}{c}{\textbf{TS}} & \multicolumn{2}{c}{\textbf{Py}} & \multicolumn{2}{c}{\textbf{JS}} & \multicolumn{2}{c}{\textbf{Java}} & \multicolumn{2}{c}{\textbf{C\#}} & \multicolumn{2}{c}{\textbf{PHP}} & \multicolumn{2}{c}{\textbf{C++}} \\
\cmidrule(lr){2-3}\cmidrule(lr){4-5}\cmidrule(lr){6-7}\cmidrule(lr){8-9}\cmidrule(lr){10-11}\cmidrule(lr){12-13}\cmidrule(lr){14-15}
\textbf{\llm} & \textbf{LC} & \textbf{CLBG} & \textbf{LC} & \textbf{CLBG} & \textbf{LC} & \textbf{CLBG} & \textbf{LC} & \textbf{CLBG} & \textbf{LC} & \textbf{CLBG} & \textbf{LC} & \textbf{CLBG} & \textbf{LC} & \textbf{CLBG} \\
\midrule
\code{gpt-6-astra} & \textbf{62} & \textbf{77} & 62 & 61 & \textbf{63} & 66 & \textbf{62} & 69 & \textbf{61} & 69 & 55 & 71 & \textbf{67} & \textbf{85} \\
\code{claude-opus-5-5} & 58 & 60 & \textbf{63} & 68 & 57 & 51 & 53 & 69 & 60 & 65 & \textbf{68} & \textbf{77} & 51 & 49 \\
\code{gpt-5.6-sol} & 55 & 64 & 62 & 55 & 50 & 51 & 57 & 56 & 59 & 60 & 56 & 51 & 51 & 72 \\
\code{claude-sonnet-5-5} & 56 & 35 & 61 & 50 & 55 & 44 & 49 & 42 & 53 & 39 & 59 & 48 & 46 & 51 \\
\code{claude-fable-5-1} & 59 & 59 & 47 & \textbf{80} & 57 & 53 & 52 & \textbf{77} & 46 & \textbf{77} & 56 & 56 & 52 & 50 \\
\code{glm-5.3-flash} & 56 & 51 & 51 & 78 & 51 & 69 & 53 & 42 & 54 & -- & 55 & -- & 49 & 54 \\
\code{gemini-3.8-flash} & 52 & 49 & 41 & 56 & 52 & 58 & 56 & 52 & 49 & 64 & 49 & 56 & 54 & 60 \\
\code{gpt-5.5} & 44 & 63 & 51 & 43 & 46 & \textbf{70} & 51 & 66 & 52 & 43 & 50 & 60 & 49 & 69 \\
\code{gpt-5.4-mini} & 48 & 46 & 40 & 13 & 48 & -- & 54 & 34 & 49 & 37 & 41 & -- & 52 & 18 \\
\code{gpt-5.6-luna} & 38 & 14 & 55 & 46 & 44 & 30 & 48 & 20 & 52 & 39 & 44 & 16 & 43 & 22 \\
\code{deepseek-v4.1-flash} & 43 & 51 & 39 & 40 & 49 & 41 & 40 & 29 & 41 & 41 & 44 & 42 & 47 & 37 \\
\code{claude-opus-4-8} & 42 & 25 & 48 & 36 & 39 & 30 & 41 & 38 & 35 & 23 & 29 & 25 & 43 & 32 \\
\code{grok-4.6} & 38 & 57 & 30 & 25 & 39 & 37 & 36 & 55 & 40 & 42 & 44 & 47 & 46 & 50 \\
\bottomrule
\end{tabular}
\end{table}

\begin{table}[H]
\centering
\footnotesize
\caption{\Energy win probability (\%) on \leetcodeComp (LC) and \clbgComp (CLBG), sorted in a descending order of \latency win probability on LC. The highest win probability in each column is in bold. A dash marks an \llm with fewer than 3~solved problems in the corresponding language.}
\label{tab:rq1-languages-energy}
\begin{tabular}{l rr rr rr rr rr rr rr}
\toprule
 & \multicolumn{2}{c}{\textbf{TS}} & \multicolumn{2}{c}{\textbf{Py}} & \multicolumn{2}{c}{\textbf{JS}} & \multicolumn{2}{c}{\textbf{Java}} & \multicolumn{2}{c}{\textbf{C\#}} & \multicolumn{2}{c}{\textbf{PHP}} & \multicolumn{2}{c}{\textbf{C++}} \\
\cmidrule(lr){2-3}\cmidrule(lr){4-5}\cmidrule(lr){6-7}\cmidrule(lr){8-9}\cmidrule(lr){10-11}\cmidrule(lr){12-13}\cmidrule(lr){14-15}
\textbf{\llm} & \textbf{LC} & \textbf{CLBG} & \textbf{LC} & \textbf{CLBG} & \textbf{LC} & \textbf{CLBG} & \textbf{LC} & \textbf{CLBG} & \textbf{LC} & \textbf{CLBG} & \textbf{LC} & \textbf{CLBG} & \textbf{LC} & \textbf{CLBG} \\
\midrule
\code{gpt-6-astra} & \textbf{61} & \textbf{76} & \textbf{64} & 61 & \textbf{62} & 66 & \textbf{62} & 70 & \textbf{64} & \textbf{70} & 50 & 73 & \textbf{67} & \textbf{83} \\
\code{claude-opus-5-5} & 58 & 56 & 63 & 67 & 57 & 50 & 52 & 69 & 53 & 52 & \textbf{71} & \textbf{77} & 52 & 49 \\
\code{gpt-5.6-sol} & 53 & 66 & 49 & 48 & 50 & 51 & 57 & 57 & 45 & 58 & 46 & 52 & 51 & 72 \\
\code{claude-sonnet-5-5} & 56 & 50 & 62 & 54 & 55 & 45 & 48 & 42 & 48 & 66 & 63 & 51 & 46 & 50 \\
\code{claude-fable-5-1} & 57 & 57 & 49 & \textbf{79} & 56 & 54 & 51 & \textbf{77} & 48 & 57 & 49 & 56 & 51 & 52 \\
\code{glm-5.3-flash} & 55 & 51 & 51 & 77 & 51 & \textbf{70} & 52 & 40 & 53 & -- & 52 & -- & 50 & 57 \\
\code{gemini-3.8-flash} & 51 & 46 & 57 & 55 & 51 & 57 & 56 & 50 & 63 & 49 & 58 & 56 & 54 & 59 \\
\code{gpt-5.5} & 45 & 64 & 50 & 43 & 46 & 68 & 52 & 66 & 50 & 43 & 46 & 58 & 48 & 64 \\
\code{gpt-5.4-mini} & 48 & 41 & 43 & 11 & 48 & -- & 53 & 37 & 50 & 39 & 35 & -- & 52 & 17 \\
\code{gpt-5.6-luna} & 39 & 16 & 47 & 46 & 45 & 28 & 50 & 20 & 38 & 38 & 51 & 21 & 44 & 23 \\
\code{deepseek-v4.1-flash} & 43 & 49 & 45 & 40 & 50 & 41 & 40 & 28 & 45 & 35 & 43 & 41 & 46 & 38 \\
\code{claude-opus-4-8} & 44 & 22 & 35 & 41 & 40 & 32 & 40 & 43 & 53 & 30 & 44 & 23 & 44 & 32 \\
\code{grok-4.6} & 39 & 54 & 34 & 27 & 40 & 38 & 36 & 51 & 40 & 62 & 42 & 43 & 46 & 55 \\
\bottomrule
\end{tabular}
\end{table}

\begin{table}[H]
\centering
\footnotesize
\caption{\Memory win probability (\%) on \leetcodeComp (LC) and \clbgComp (CLBG), sorted in a descending order of \latency win probability on LC. The highest win probability in each column is in bold. A dash marks an \llm with fewer than 3~solved problems in the corresponding language.}
\label{tab:rq1-languages-memory}
\begin{tabular}{l rr rr rr rr rr rr rr}
\toprule
 & \multicolumn{2}{c}{\textbf{TS}} & \multicolumn{2}{c}{\textbf{Py}} & \multicolumn{2}{c}{\textbf{JS}} & \multicolumn{2}{c}{\textbf{Java}} & \multicolumn{2}{c}{\textbf{C\#}} & \multicolumn{2}{c}{\textbf{PHP}} & \multicolumn{2}{c}{\textbf{C++}} \\
\cmidrule(lr){2-3}\cmidrule(lr){4-5}\cmidrule(lr){6-7}\cmidrule(lr){8-9}\cmidrule(lr){10-11}\cmidrule(lr){12-13}\cmidrule(lr){14-15}
\textbf{\llm} & \textbf{LC} & \textbf{CLBG} & \textbf{LC} & \textbf{CLBG} & \textbf{LC} & \textbf{CLBG} & \textbf{LC} & \textbf{CLBG} & \textbf{LC} & \textbf{CLBG} & \textbf{LC} & \textbf{CLBG} & \textbf{LC} & \textbf{CLBG} \\
\midrule
\code{gpt-6-astra} & 52 & \textbf{77} & 50 & 58 & 50 & \textbf{71} & 51 & \textbf{60} & 50 & \textbf{61} & 50 & \textbf{68} & 48 & 56 \\
\code{claude-opus-5-5} & 52 & 45 & \textbf{52} & 24 & 49 & 49 & 50 & 46 & \textbf{51} & 48 & \textbf{55} & 46 & 51 & 37 \\
\code{gpt-5.6-sol} & 48 & 50 & 50 & 54 & 48 & 57 & 51 & 46 & 50 & 55 & 50 & 39 & \textbf{52} & 42 \\
\code{claude-sonnet-5-5} & 50 & 57 & \textbf{52} & 25 & 50 & 51 & 49 & 55 & 50 & 43 & 54 & 43 & 50 & 50 \\
\code{claude-fable-5-1} & \textbf{53} & 53 & 50 & 40 & 50 & 46 & 50 & 50 & 50 & 47 & 54 & 44 & 48 & 54 \\
\code{glm-5.3-flash} & 48 & 46 & 47 & 41 & \textbf{52} & 52 & 44 & 50 & 50 & -- & 50 & -- & 49 & 45 \\
\code{gemini-3.8-flash} & 51 & 52 & 50 & 63 & \textbf{52} & 51 & 48 & 44 & 50 & 50 & 49 & 52 & \textbf{52} & 38 \\
\code{gpt-5.5} & 49 & 41 & 50 & 55 & 48 & 44 & 50 & 43 & 50 & 47 & 47 & 45 & 49 & 43 \\
\code{gpt-5.4-mini} & 47 & 52 & 50 & 41 & 49 & -- & 50 & 48 & 49 & 48 & 45 & -- & 51 & 36 \\
\code{gpt-5.6-luna} & 47 & 25 & 50 & 66 & 49 & 40 & \textbf{55} & 56 & 50 & 52 & 47 & 58 & 48 & 65 \\
\code{deepseek-v4.1-flash} & 51 & 50 & 50 & \textbf{72} & 49 & 30 & 53 & 52 & 49 & 47 & 49 & 37 & 51 & \textbf{76} \\
\code{claude-opus-4-8} & 50 & 43 & 48 & 61 & 51 & 61 & 50 & 59 & 50 & 46 & 52 & 58 & 51 & 43 \\
\code{grok-4.6} & 52 & 59 & 51 & 49 & 51 & 48 & 50 & 42 & 50 & 55 & 48 & 60 & 51 & 66 \\
\bottomrule
\end{tabular}
\end{table}

\begin{table}[H]
\centering
\scriptsize
\setlength{\tabcolsep}{3pt}
\caption{\llms on TypeScript, ranked by win probability (\%) in \latency, \energy, and \memory.}
\label{tab:bt-typescript}
\subfloat[\leetcodeComp\label{tab:bt-leetcode-99-perf-typescript}]{\begin{tabular}{@{}rl rl rl@{}}
\toprule
\multicolumn{2}{c}{\textbf{Latency}} & \multicolumn{2}{c}{\textbf{Energy}} & \multicolumn{2}{c}{\textbf{Peak memory}} \\
\cmidrule(lr){1-2}\cmidrule(lr){3-4}\cmidrule(l){5-6}
\textbf{62} & \textbf{\code{gpt-6-astra}} & \textbf{61} & \textbf{\code{gpt-6-astra}} & \textbf{53} & \textbf{\code{claude-fable-5-1}} \\
\textbf{59} & \textbf{\code{claude-fable-5-1}} & \textbf{58} & \textbf{\code{claude-opus-5-5}} & \textbf{52} & \textbf{\code{grok-4.6}} \\
\textbf{58} & \textbf{\code{claude-opus-5-5}} & \textbf{57} & \textbf{\code{claude-fable-5-1}} & \textbf{52} & \textbf{\code{claude-opus-5-5}} \\
\textbf{56} & \textbf{\code{claude-sonnet-5-5}} & \textbf{56} & \textbf{\code{claude-sonnet-5-5}} & \textbf{52} & \textbf{\code{gpt-6-astra}} \\
\textbf{56} & \textbf{\code{glm-5.3-flash}} & \textbf{55} & \textbf{\code{glm-5.3-flash}} & \textbf{51} & \textbf{\code{gemini-3.8-flash}} \\
\textbf{55} & \textbf{\code{gpt-5.6-sol}} & \textbf{53} & \textbf{\code{gpt-5.6-sol}} & \textbf{51} & \textbf{\code{deepseek-v4.1-flash}} \\
\textbf{52} & \textbf{\code{gemini-3.8-flash}} & \textbf{51} & \textbf{\code{gemini-3.8-flash}} & \textbf{50} & \textbf{\code{claude-opus-4-8}} \\
\textbf{48} & \textbf{\code{gpt-5.4-mini}} & \textbf{48} & \textbf{\code{gpt-5.4-mini}} & \textbf{50} & \textbf{\code{claude-sonnet-5-5}} \\
44 & \code{gpt-5.5} & 45 & \code{gpt-5.5} & \textbf{49} & \textbf{\code{gpt-5.5}} \\
43 & \code{deepseek-v4.1-flash} & 44 & \code{claude-opus-4-8} & \textbf{48} & \textbf{\code{gpt-5.6-sol}} \\
42 & \code{claude-opus-4-8} & 43 & \code{deepseek-v4.1-flash} & \textbf{48} & \textbf{\code{glm-5.3-flash}} \\
38 & \code{grok-4.6} & 39 & \code{grok-4.6} & \textbf{47} & \textbf{\code{gpt-5.4-mini}} \\
38 & \code{gpt-5.6-luna} & 39 & \code{gpt-5.6-luna} & \textbf{47} & \textbf{\code{gpt-5.6-luna}} \\
\bottomrule
\end{tabular}}\hfil
\subfloat[\clbgComp\label{tab:bt-clbg-10-typescript}]{\begin{tabular}{@{}rl rl rl@{}}
\toprule
\multicolumn{2}{c}{\textbf{Latency}} & \multicolumn{2}{c}{\textbf{Energy}} & \multicolumn{2}{c}{\textbf{Peak memory}} \\
\cmidrule(lr){1-2}\cmidrule(lr){3-4}\cmidrule(l){5-6}
\textbf{77} & \textbf{\code{gpt-6-astra}} & \textbf{76} & \textbf{\code{gpt-6-astra}} & \textbf{77} & \textbf{\code{gpt-6-astra}} \\
\textbf{64} & \textbf{\code{gpt-5.6-sol}} & \textbf{66} & \textbf{\code{gpt-5.6-sol}} & \textbf{59} & \textbf{\code{grok-4.6}} \\
\textbf{63} & \textbf{\code{gpt-5.5}} & \textbf{64} & \textbf{\code{gpt-5.5}} & \textbf{57} & \textbf{\code{claude-sonnet-5-5}} \\
\textbf{60} & \textbf{\code{claude-opus-5-5}} & \textbf{57} & \textbf{\code{claude-fable-5-1}} & \textbf{53} & \textbf{\code{claude-fable-5-1}} \\
\textbf{59} & \textbf{\code{claude-fable-5-1}} & \textbf{56} & \textbf{\code{claude-opus-5-5}} & \textbf{52} & \textbf{\code{gpt-5.4-mini}} \\
\textbf{57} & \textbf{\code{grok-4.6}} & \textbf{54} & \textbf{\code{grok-4.6}} & \textbf{52} & \textbf{\code{gemini-3.8-flash}} \\
\textbf{51} & \textbf{\code{deepseek-v4.1-flash}} & \textbf{51} & \textbf{\code{glm-5.3-flash}} & \textbf{50} & \textbf{\code{gpt-5.6-sol}} \\
\textbf{51} & \textbf{\code{glm-5.3-flash}} & \textbf{50} & \textbf{\code{claude-sonnet-5-5}} & \textbf{50} & \textbf{\code{deepseek-v4.1-flash}} \\
\textbf{49} & \textbf{\code{gemini-3.8-flash}} & \textbf{49} & \textbf{\code{deepseek-v4.1-flash}} & \textbf{46} & \textbf{\code{glm-5.3-flash}} \\
\textbf{46} & \textbf{\code{gpt-5.4-mini}} & \textbf{46} & \textbf{\code{gemini-3.8-flash}} & \textbf{45} & \textbf{\code{claude-opus-5-5}} \\
35 & \code{claude-sonnet-5-5} & \textbf{41} & \textbf{\code{gpt-5.4-mini}} & \textbf{43} & \textbf{\code{claude-opus-4-8}} \\
\textbf{25} & \textbf{\code{claude-opus-4-8}} & 22 & \code{claude-opus-4-8} & 41 & \code{gpt-5.5} \\
14 & \code{gpt-5.6-luna} & 16 & \code{gpt-5.6-luna} & 25 & \code{gpt-5.6-luna} \\
\bottomrule
\end{tabular}}
\end{table}

\begin{table}[H]
\centering
\scriptsize
\setlength{\tabcolsep}{3pt}
\caption{\llms on Python, ranked by win probability (\%) in \latency, \energy, and \memory.}
\label{tab:bt-python3}
\subfloat[\leetcodeComp\label{tab:bt-leetcode-99-perf-python3}]{\begin{tabular}{@{}rl rl rl@{}}
\toprule
\multicolumn{2}{c}{\textbf{Latency}} & \multicolumn{2}{c}{\textbf{Energy}} & \multicolumn{2}{c}{\textbf{Peak memory}} \\
\cmidrule(lr){1-2}\cmidrule(lr){3-4}\cmidrule(l){5-6}
\textbf{63} & \textbf{\code{claude-opus-5-5}} & \textbf{64} & \textbf{\code{gpt-6-astra}} & \textbf{52} & \textbf{\code{claude-opus-5-5}} \\
\textbf{62} & \textbf{\code{gpt-6-astra}} & \textbf{63} & \textbf{\code{claude-opus-5-5}} & \textbf{52} & \textbf{\code{claude-sonnet-5-5}} \\
\textbf{62} & \textbf{\code{gpt-5.6-sol}} & \textbf{62} & \textbf{\code{claude-sonnet-5-5}} & \textbf{51} & \textbf{\code{grok-4.6}} \\
\textbf{61} & \textbf{\code{claude-sonnet-5-5}} & \textbf{57} & \textbf{\code{gemini-3.8-flash}} & \textbf{50} & \textbf{\code{gpt-6-astra}} \\
\textbf{55} & \textbf{\code{gpt-5.6-luna}} & \textbf{51} & \textbf{\code{glm-5.3-flash}} & \textbf{50} & \textbf{\code{deepseek-v4.1-flash}} \\
\textbf{51} & \textbf{\code{gpt-5.5}} & 50 & \code{gpt-5.5} & \textbf{50} & \textbf{\code{gpt-5.4-mini}} \\
\textbf{51} & \textbf{\code{glm-5.3-flash}} & 49 & \code{gpt-5.6-sol} & \textbf{50} & \textbf{\code{gpt-5.6-luna}} \\
48 & \code{claude-opus-4-8} & 49 & \code{claude-fable-5-1} & \textbf{50} & \textbf{\code{gemini-3.8-flash}} \\
47 & \code{claude-fable-5-1} & 47 & \code{gpt-5.6-luna} & \textbf{50} & \textbf{\code{gpt-5.6-sol}} \\
41 & \code{gemini-3.8-flash} & 45 & \code{deepseek-v4.1-flash} & \textbf{50} & \textbf{\code{gpt-5.5}} \\
40 & \code{gpt-5.4-mini} & 43 & \code{gpt-5.4-mini} & \textbf{50} & \textbf{\code{claude-fable-5-1}} \\
39 & \code{deepseek-v4.1-flash} & 35 & \code{claude-opus-4-8} & \textbf{48} & \textbf{\code{claude-opus-4-8}} \\
30 & \code{grok-4.6} & 34 & \code{grok-4.6} & \textbf{47} & \textbf{\code{glm-5.3-flash}} \\
\bottomrule
\end{tabular}}\hfil
\subfloat[\clbgComp\label{tab:bt-clbg-10-python3}]{\begin{tabular}{@{}rl rl rl@{}}
\toprule
\multicolumn{2}{c}{\textbf{Latency}} & \multicolumn{2}{c}{\textbf{Energy}} & \multicolumn{2}{c}{\textbf{Peak memory}} \\
\cmidrule(lr){1-2}\cmidrule(lr){3-4}\cmidrule(l){5-6}
\textbf{80} & \textbf{\code{claude-fable-5-1}} & \textbf{79} & \textbf{\code{claude-fable-5-1}} & \textbf{72} & \textbf{\code{deepseek-v4.1-flash}} \\
\textbf{78} & \textbf{\code{glm-5.3-flash}} & \textbf{77} & \textbf{\code{glm-5.3-flash}} & \textbf{66} & \textbf{\code{gpt-5.6-luna}} \\
\textbf{68} & \textbf{\code{claude-opus-5-5}} & \textbf{67} & \textbf{\code{claude-opus-5-5}} & \textbf{63} & \textbf{\code{gemini-3.8-flash}} \\
\textbf{61} & \textbf{\code{gpt-6-astra}} & \textbf{61} & \textbf{\code{gpt-6-astra}} & \textbf{61} & \textbf{\code{claude-opus-4-8}} \\
\textbf{56} & \textbf{\code{gemini-3.8-flash}} & \textbf{55} & \textbf{\code{gemini-3.8-flash}} & \textbf{58} & \textbf{\code{gpt-6-astra}} \\
\textbf{55} & \textbf{\code{gpt-5.6-sol}} & \textbf{54} & \textbf{\code{claude-sonnet-5-5}} & \textbf{55} & \textbf{\code{gpt-5.5}} \\
\textbf{50} & \textbf{\code{claude-sonnet-5-5}} & \textbf{48} & \textbf{\code{gpt-5.6-sol}} & \textbf{54} & \textbf{\code{gpt-5.6-sol}} \\
\textbf{46} & \textbf{\code{gpt-5.6-luna}} & \textbf{46} & \textbf{\code{gpt-5.6-luna}} & \textbf{49} & \textbf{\code{grok-4.6}} \\
43 & \code{gpt-5.5} & \textbf{43} & \textbf{\code{gpt-5.5}} & \textbf{41} & \textbf{\code{gpt-5.4-mini}} \\
40 & \code{deepseek-v4.1-flash} & \textbf{41} & \textbf{\code{claude-opus-4-8}} & \textbf{41} & \textbf{\code{glm-5.3-flash}} \\
\textbf{36} & \textbf{\code{claude-opus-4-8}} & 40 & \code{deepseek-v4.1-flash} & \textbf{40} & \textbf{\code{claude-fable-5-1}} \\
25 & \code{grok-4.6} & 27 & \code{grok-4.6} & \textbf{25} & \textbf{\code{claude-sonnet-5-5}} \\
13 & \code{gpt-5.4-mini} & 11 & \code{gpt-5.4-mini} & \textbf{24} & \textbf{\code{claude-opus-5-5}} \\
\bottomrule
\end{tabular}}
\end{table}

\begin{table}[H]
\centering
\scriptsize
\setlength{\tabcolsep}{3pt}
\caption{\llms on JavaScript, ranked by win probability (\%) in \latency, \energy, and \memory.}
\label{tab:bt-javascript}
\subfloat[\leetcodeComp\label{tab:bt-leetcode-99-perf-javascript}]{\begin{tabular}{@{}rl rl rl@{}}
\toprule
\multicolumn{2}{c}{\textbf{Latency}} & \multicolumn{2}{c}{\textbf{Energy}} & \multicolumn{2}{c}{\textbf{Peak memory}} \\
\cmidrule(lr){1-2}\cmidrule(lr){3-4}\cmidrule(l){5-6}
\textbf{63} & \textbf{\code{gpt-6-astra}} & \textbf{62} & \textbf{\code{gpt-6-astra}} & \textbf{52} & \textbf{\code{glm-5.3-flash}} \\
\textbf{57} & \textbf{\code{claude-opus-5-5}} & \textbf{57} & \textbf{\code{claude-opus-5-5}} & \textbf{52} & \textbf{\code{gemini-3.8-flash}} \\
\textbf{57} & \textbf{\code{claude-fable-5-1}} & \textbf{56} & \textbf{\code{claude-fable-5-1}} & \textbf{51} & \textbf{\code{claude-opus-4-8}} \\
\textbf{55} & \textbf{\code{claude-sonnet-5-5}} & \textbf{55} & \textbf{\code{claude-sonnet-5-5}} & \textbf{51} & \textbf{\code{grok-4.6}} \\
\textbf{52} & \textbf{\code{gemini-3.8-flash}} & \textbf{51} & \textbf{\code{glm-5.3-flash}} & \textbf{50} & \textbf{\code{gpt-6-astra}} \\
\textbf{51} & \textbf{\code{glm-5.3-flash}} & \textbf{51} & \textbf{\code{gemini-3.8-flash}} & \textbf{50} & \textbf{\code{claude-sonnet-5-5}} \\
50 & \code{gpt-5.6-sol} & \textbf{50} & \textbf{\code{deepseek-v4.1-flash}} & \textbf{50} & \textbf{\code{claude-fable-5-1}} \\
\textbf{49} & \textbf{\code{deepseek-v4.1-flash}} & 50 & \code{gpt-5.6-sol} & \textbf{49} & \textbf{\code{deepseek-v4.1-flash}} \\
\textbf{48} & \textbf{\code{gpt-5.4-mini}} & \textbf{48} & \textbf{\code{gpt-5.4-mini}} & \textbf{49} & \textbf{\code{claude-opus-5-5}} \\
46 & \code{gpt-5.5} & 46 & \code{gpt-5.5} & \textbf{49} & \textbf{\code{gpt-5.6-luna}} \\
44 & \code{gpt-5.6-luna} & 45 & \code{gpt-5.6-luna} & \textbf{49} & \textbf{\code{gpt-5.4-mini}} \\
39 & \code{grok-4.6} & 40 & \code{grok-4.6} & \textbf{48} & \textbf{\code{gpt-5.5}} \\
39 & \code{claude-opus-4-8} & 40 & \code{claude-opus-4-8} & \textbf{48} & \textbf{\code{gpt-5.6-sol}} \\
\bottomrule
\end{tabular}}\hfil
\subfloat[\clbgComp\label{tab:bt-clbg-10-javascript}]{\begin{tabular}{@{}rl rl rl@{}}
\toprule
\multicolumn{2}{c}{\textbf{Latency}} & \multicolumn{2}{c}{\textbf{Energy}} & \multicolumn{2}{c}{\textbf{Peak memory}} \\
\cmidrule(lr){1-2}\cmidrule(lr){3-4}\cmidrule(l){5-6}
\textbf{70} & \textbf{\code{gpt-5.5}} & \textbf{70} & \textbf{\code{glm-5.3-flash}} & \textbf{71} & \textbf{\code{gpt-6-astra}} \\
\textbf{69} & \textbf{\code{glm-5.3-flash}} & \textbf{68} & \textbf{\code{gpt-5.5}} & \textbf{61} & \textbf{\code{claude-opus-4-8}} \\
\textbf{66} & \textbf{\code{gpt-6-astra}} & \textbf{66} & \textbf{\code{gpt-6-astra}} & \textbf{57} & \textbf{\code{gpt-5.6-sol}} \\
\textbf{58} & \textbf{\code{gemini-3.8-flash}} & \textbf{57} & \textbf{\code{gemini-3.8-flash}} & \textbf{52} & \textbf{\code{glm-5.3-flash}} \\
\textbf{53} & \textbf{\code{claude-fable-5-1}} & \textbf{54} & \textbf{\code{claude-fable-5-1}} & \textbf{51} & \textbf{\code{gemini-3.8-flash}} \\
\textbf{51} & \textbf{\code{gpt-5.6-sol}} & \textbf{51} & \textbf{\code{gpt-5.6-sol}} & \textbf{51} & \textbf{\code{claude-sonnet-5-5}} \\
\textbf{51} & \textbf{\code{claude-opus-5-5}} & \textbf{50} & \textbf{\code{claude-opus-5-5}} & \textbf{49} & \textbf{\code{claude-opus-5-5}} \\
\textbf{44} & \textbf{\code{claude-sonnet-5-5}} & \textbf{45} & \textbf{\code{claude-sonnet-5-5}} & 48 & \code{grok-4.6} \\
\textbf{41} & \textbf{\code{deepseek-v4.1-flash}} & \textbf{41} & \textbf{\code{deepseek-v4.1-flash}} & \textbf{46} & \textbf{\code{claude-fable-5-1}} \\
\textbf{37} & \textbf{\code{grok-4.6}} & \textbf{38} & \textbf{\code{grok-4.6}} & \textbf{44} & \textbf{\code{gpt-5.5}} \\
\textbf{30} & \textbf{\code{gpt-5.6-luna}} & \textbf{32} & \textbf{\code{claude-opus-4-8}} & \textbf{40} & \textbf{\code{gpt-5.6-luna}} \\
\textbf{30} & \textbf{\code{claude-opus-4-8}} & \textbf{28} & \textbf{\code{gpt-5.6-luna}} & 30 & \code{deepseek-v4.1-flash} \\
-- & \code{gpt-5.4-mini} & -- & \code{gpt-5.4-mini} & -- & \code{gpt-5.4-mini} \\
\bottomrule
\end{tabular}}
\end{table}

\begin{table}[H]
\centering
\scriptsize
\setlength{\tabcolsep}{3pt}
\caption{\llms on Java, ranked by win probability (\%) in \latency, \energy, and \memory.}
\label{tab:bt-java}
\subfloat[\leetcodeComp\label{tab:bt-leetcode-99-perf-java}]{\begin{tabular}{@{}rl rl rl@{}}
\toprule
\multicolumn{2}{c}{\textbf{Latency}} & \multicolumn{2}{c}{\textbf{Energy}} & \multicolumn{2}{c}{\textbf{Peak memory}} \\
\cmidrule(lr){1-2}\cmidrule(lr){3-4}\cmidrule(l){5-6}
\textbf{62} & \textbf{\code{gpt-6-astra}} & \textbf{62} & \textbf{\code{gpt-6-astra}} & \textbf{55} & \textbf{\code{gpt-5.6-luna}} \\
\textbf{57} & \textbf{\code{gpt-5.6-sol}} & \textbf{57} & \textbf{\code{gpt-5.6-sol}} & \textbf{53} & \textbf{\code{deepseek-v4.1-flash}} \\
\textbf{56} & \textbf{\code{gemini-3.8-flash}} & \textbf{56} & \textbf{\code{gemini-3.8-flash}} & \textbf{51} & \textbf{\code{gpt-6-astra}} \\
\textbf{54} & \textbf{\code{gpt-5.4-mini}} & \textbf{53} & \textbf{\code{gpt-5.4-mini}} & \textbf{51} & \textbf{\code{gpt-5.6-sol}} \\
\textbf{53} & \textbf{\code{claude-opus-5-5}} & \textbf{52} & \textbf{\code{claude-opus-5-5}} & \textbf{50} & \textbf{\code{claude-opus-4-8}} \\
\textbf{53} & \textbf{\code{glm-5.3-flash}} & \textbf{52} & \textbf{\code{glm-5.3-flash}} & \textbf{50} & \textbf{\code{claude-fable-5-1}} \\
\textbf{52} & \textbf{\code{claude-fable-5-1}} & \textbf{52} & \textbf{\code{gpt-5.5}} & \textbf{50} & \textbf{\code{grok-4.6}} \\
\textbf{51} & \textbf{\code{gpt-5.5}} & \textbf{51} & \textbf{\code{claude-fable-5-1}} & \textbf{50} & \textbf{\code{claude-opus-5-5}} \\
\textbf{49} & \textbf{\code{claude-sonnet-5-5}} & \textbf{50} & \textbf{\code{gpt-5.6-luna}} & \textbf{50} & \textbf{\code{gpt-5.4-mini}} \\
\textbf{48} & \textbf{\code{gpt-5.6-luna}} & 48 & \code{claude-sonnet-5-5} & \textbf{50} & \textbf{\code{gpt-5.5}} \\
41 & \code{claude-opus-4-8} & 40 & \code{deepseek-v4.1-flash} & \textbf{49} & \textbf{\code{claude-sonnet-5-5}} \\
40 & \code{deepseek-v4.1-flash} & 40 & \code{claude-opus-4-8} & \textbf{48} & \textbf{\code{gemini-3.8-flash}} \\
36 & \code{grok-4.6} & 36 & \code{grok-4.6} & \textbf{44} & \textbf{\code{glm-5.3-flash}} \\
\bottomrule
\end{tabular}}\hfil
\subfloat[\clbgComp\label{tab:bt-clbg-10-java}]{\begin{tabular}{@{}rl rl rl@{}}
\toprule
\multicolumn{2}{c}{\textbf{Latency}} & \multicolumn{2}{c}{\textbf{Energy}} & \multicolumn{2}{c}{\textbf{Peak memory}} \\
\cmidrule(lr){1-2}\cmidrule(lr){3-4}\cmidrule(l){5-6}
\textbf{77} & \textbf{\code{claude-fable-5-1}} & \textbf{77} & \textbf{\code{claude-fable-5-1}} & \textbf{60} & \textbf{\code{gpt-6-astra}} \\
\textbf{69} & \textbf{\code{claude-opus-5-5}} & \textbf{70} & \textbf{\code{gpt-6-astra}} & \textbf{59} & \textbf{\code{claude-opus-4-8}} \\
\textbf{69} & \textbf{\code{gpt-6-astra}} & \textbf{69} & \textbf{\code{claude-opus-5-5}} & \textbf{56} & \textbf{\code{gpt-5.6-luna}} \\
\textbf{66} & \textbf{\code{gpt-5.5}} & \textbf{66} & \textbf{\code{gpt-5.5}} & \textbf{55} & \textbf{\code{claude-sonnet-5-5}} \\
\textbf{56} & \textbf{\code{gpt-5.6-sol}} & \textbf{57} & \textbf{\code{gpt-5.6-sol}} & \textbf{52} & \textbf{\code{deepseek-v4.1-flash}} \\
\textbf{55} & \textbf{\code{grok-4.6}} & \textbf{51} & \textbf{\code{grok-4.6}} & \textbf{50} & \textbf{\code{glm-5.3-flash}} \\
\textbf{52} & \textbf{\code{gemini-3.8-flash}} & \textbf{50} & \textbf{\code{gemini-3.8-flash}} & \textbf{50} & \textbf{\code{claude-fable-5-1}} \\
\textbf{42} & \textbf{\code{glm-5.3-flash}} & \textbf{43} & \textbf{\code{claude-opus-4-8}} & \textbf{48} & \textbf{\code{gpt-5.4-mini}} \\
42 & \code{claude-sonnet-5-5} & 42 & \code{claude-sonnet-5-5} & \textbf{46} & \textbf{\code{claude-opus-5-5}} \\
\textbf{38} & \textbf{\code{claude-opus-4-8}} & \textbf{40} & \textbf{\code{glm-5.3-flash}} & \textbf{46} & \textbf{\code{gpt-5.6-sol}} \\
\textbf{34} & \textbf{\code{gpt-5.4-mini}} & \textbf{37} & \textbf{\code{gpt-5.4-mini}} & \textbf{44} & \textbf{\code{gemini-3.8-flash}} \\
29 & \code{deepseek-v4.1-flash} & 28 & \code{deepseek-v4.1-flash} & \textbf{43} & \textbf{\code{gpt-5.5}} \\
20 & \code{gpt-5.6-luna} & 20 & \code{gpt-5.6-luna} & 42 & \code{grok-4.6} \\
\bottomrule
\end{tabular}}
\end{table}

\begin{table}[H]
\centering
\scriptsize
\setlength{\tabcolsep}{3pt}
\caption{\llms on C\#, ranked by win probability (\%) in \latency, \energy, and \memory.}
\label{tab:bt-csharp}
\subfloat[\leetcodeComp\label{tab:bt-leetcode-99-perf-csharp}]{\begin{tabular}{@{}rl rl rl@{}}
\toprule
\multicolumn{2}{c}{\textbf{Latency}} & \multicolumn{2}{c}{\textbf{Energy}} & \multicolumn{2}{c}{\textbf{Peak memory}} \\
\cmidrule(lr){1-2}\cmidrule(lr){3-4}\cmidrule(l){5-6}
\textbf{61} & \textbf{\code{gpt-6-astra}} & \textbf{64} & \textbf{\code{gpt-6-astra}} & \textbf{51} & \textbf{\code{claude-opus-5-5}} \\
\textbf{60} & \textbf{\code{claude-opus-5-5}} & \textbf{63} & \textbf{\code{gemini-3.8-flash}} & \textbf{50} & \textbf{\code{claude-fable-5-1}} \\
\textbf{59} & \textbf{\code{gpt-5.6-sol}} & \textbf{53} & \textbf{\code{claude-opus-5-5}} & \textbf{50} & \textbf{\code{claude-sonnet-5-5}} \\
\textbf{54} & \textbf{\code{glm-5.3-flash}} & \textbf{53} & \textbf{\code{glm-5.3-flash}} & \textbf{50} & \textbf{\code{claude-opus-4-8}} \\
\textbf{53} & \textbf{\code{claude-sonnet-5-5}} & \textbf{53} & \textbf{\code{claude-opus-4-8}} & \textbf{50} & \textbf{\code{gpt-5.6-luna}} \\
\textbf{52} & \textbf{\code{gpt-5.6-luna}} & 50 & \code{gpt-5.5} & \textbf{50} & \textbf{\code{gpt-5.6-sol}} \\
\textbf{52} & \textbf{\code{gpt-5.5}} & 50 & \code{gpt-5.4-mini} & \textbf{50} & \textbf{\code{gpt-6-astra}} \\
\textbf{49} & \textbf{\code{gemini-3.8-flash}} & 48 & \code{claude-sonnet-5-5} & \textbf{50} & \textbf{\code{gpt-5.5}} \\
\textbf{49} & \textbf{\code{gpt-5.4-mini}} & 48 & \code{claude-fable-5-1} & \textbf{50} & \textbf{\code{glm-5.3-flash}} \\
46 & \code{claude-fable-5-1} & 45 & \code{gpt-5.6-sol} & \textbf{50} & \textbf{\code{gemini-3.8-flash}} \\
41 & \code{deepseek-v4.1-flash} & 45 & \code{deepseek-v4.1-flash} & \textbf{50} & \textbf{\code{grok-4.6}} \\
40 & \code{grok-4.6} & 40 & \code{grok-4.6} & \textbf{49} & \textbf{\code{gpt-5.4-mini}} \\
35 & \code{claude-opus-4-8} & 38 & \code{gpt-5.6-luna} & \textbf{49} & \textbf{\code{deepseek-v4.1-flash}} \\
\bottomrule
\end{tabular}}\hfil
\subfloat[\clbgComp\label{tab:bt-clbg-10-csharp}]{\begin{tabular}{@{}rl rl rl@{}}
\toprule
\multicolumn{2}{c}{\textbf{Latency}} & \multicolumn{2}{c}{\textbf{Energy}} & \multicolumn{2}{c}{\textbf{Peak memory}} \\
\cmidrule(lr){1-2}\cmidrule(lr){3-4}\cmidrule(l){5-6}
\textbf{77} & \textbf{\code{claude-fable-5-1}} & \textbf{70} & \textbf{\code{gpt-6-astra}} & \textbf{61} & \textbf{\code{gpt-6-astra}} \\
\textbf{69} & \textbf{\code{gpt-6-astra}} & \textbf{66} & \textbf{\code{claude-sonnet-5-5}} & \textbf{55} & \textbf{\code{gpt-5.6-sol}} \\
\textbf{65} & \textbf{\code{claude-opus-5-5}} & \textbf{62} & \textbf{\code{grok-4.6}} & \textbf{55} & \textbf{\code{grok-4.6}} \\
\textbf{64} & \textbf{\code{gemini-3.8-flash}} & \textbf{58} & \textbf{\code{gpt-5.6-sol}} & \textbf{52} & \textbf{\code{gpt-5.6-luna}} \\
\textbf{60} & \textbf{\code{gpt-5.6-sol}} & \textbf{57} & \textbf{\code{claude-fable-5-1}} & \textbf{50} & \textbf{\code{gemini-3.8-flash}} \\
\textbf{43} & \textbf{\code{gpt-5.5}} & \textbf{52} & \textbf{\code{claude-opus-5-5}} & \textbf{48} & \textbf{\code{claude-opus-5-5}} \\
\textbf{42} & \textbf{\code{grok-4.6}} & \textbf{49} & \textbf{\code{gemini-3.8-flash}} & \textbf{48} & \textbf{\code{gpt-5.4-mini}} \\
\textbf{41} & \textbf{\code{deepseek-v4.1-flash}} & \textbf{43} & \textbf{\code{gpt-5.5}} & \textbf{47} & \textbf{\code{claude-fable-5-1}} \\
39 & \code{claude-sonnet-5-5} & \textbf{39} & \textbf{\code{gpt-5.4-mini}} & \textbf{47} & \textbf{\code{deepseek-v4.1-flash}} \\
\textbf{39} & \textbf{\code{gpt-5.6-luna}} & \textbf{38} & \textbf{\code{gpt-5.6-luna}} & \textbf{47} & \textbf{\code{gpt-5.5}} \\
\textbf{37} & \textbf{\code{gpt-5.4-mini}} & \textbf{35} & \textbf{\code{deepseek-v4.1-flash}} & \textbf{46} & \textbf{\code{claude-opus-4-8}} \\
23 & \code{claude-opus-4-8} & 30 & \code{claude-opus-4-8} & \textbf{43} & \textbf{\code{claude-sonnet-5-5}} \\
-- & \code{glm-5.3-flash} & -- & \code{glm-5.3-flash} & -- & \code{glm-5.3-flash} \\
\bottomrule
\end{tabular}}
\end{table}

\begin{table}[H]
\centering
\scriptsize
\setlength{\tabcolsep}{3pt}
\caption{\llms on PHP, ranked by win probability (\%) in \latency, \energy, and \memory.}
\label{tab:bt-php}
\subfloat[\leetcodeComp\label{tab:bt-leetcode-99-perf-php}]{\begin{tabular}{@{}rl rl rl@{}}
\toprule
\multicolumn{2}{c}{\textbf{Latency}} & \multicolumn{2}{c}{\textbf{Energy}} & \multicolumn{2}{c}{\textbf{Peak memory}} \\
\cmidrule(lr){1-2}\cmidrule(lr){3-4}\cmidrule(l){5-6}
\textbf{68} & \textbf{\code{claude-opus-5-5}} & \textbf{71} & \textbf{\code{claude-opus-5-5}} & \textbf{55} & \textbf{\code{claude-opus-5-5}} \\
\textbf{59} & \textbf{\code{claude-sonnet-5-5}} & \textbf{63} & \textbf{\code{claude-sonnet-5-5}} & \textbf{54} & \textbf{\code{claude-fable-5-1}} \\
\textbf{56} & \textbf{\code{claude-fable-5-1}} & \textbf{58} & \textbf{\code{gemini-3.8-flash}} & \textbf{54} & \textbf{\code{claude-sonnet-5-5}} \\
\textbf{56} & \textbf{\code{gpt-5.6-sol}} & 52 & \code{glm-5.3-flash} & \textbf{52} & \textbf{\code{claude-opus-4-8}} \\
\textbf{55} & \textbf{\code{gpt-6-astra}} & 51 & \code{gpt-5.6-luna} & \textbf{50} & \textbf{\code{gpt-6-astra}} \\
\textbf{55} & \textbf{\code{glm-5.3-flash}} & 50 & \code{gpt-6-astra} & \textbf{50} & \textbf{\code{gpt-5.6-sol}} \\
50 & \code{gpt-5.5} & 49 & \code{claude-fable-5-1} & 50 & \code{glm-5.3-flash} \\
49 & \code{gemini-3.8-flash} & 46 & \code{gpt-5.6-sol} & 49 & \code{deepseek-v4.1-flash} \\
44 & \code{deepseek-v4.1-flash} & 46 & \code{gpt-5.5} & \textbf{49} & \textbf{\code{gemini-3.8-flash}} \\
44 & \code{gpt-5.6-luna} & 44 & \code{claude-opus-4-8} & \textbf{48} & \textbf{\code{grok-4.6}} \\
44 & \code{grok-4.6} & 43 & \code{deepseek-v4.1-flash} & 47 & \code{gpt-5.6-luna} \\
41 & \code{gpt-5.4-mini} & 42 & \code{grok-4.6} & 47 & \code{gpt-5.5} \\
29 & \code{claude-opus-4-8} & 35 & \code{gpt-5.4-mini} & 45 & \code{gpt-5.4-mini} \\
\bottomrule
\end{tabular}}\hfil
\subfloat[\clbgComp\label{tab:bt-clbg-10-php}]{\begin{tabular}{@{}rl rl rl@{}}
\toprule
\multicolumn{2}{c}{\textbf{Latency}} & \multicolumn{2}{c}{\textbf{Energy}} & \multicolumn{2}{c}{\textbf{Peak memory}} \\
\cmidrule(lr){1-2}\cmidrule(lr){3-4}\cmidrule(l){5-6}
\textbf{77} & \textbf{\code{claude-opus-5-5}} & \textbf{77} & \textbf{\code{claude-opus-5-5}} & \textbf{68} & \textbf{\code{gpt-6-astra}} \\
\textbf{71} & \textbf{\code{gpt-6-astra}} & \textbf{73} & \textbf{\code{gpt-6-astra}} & \textbf{60} & \textbf{\code{grok-4.6}} \\
\textbf{60} & \textbf{\code{gpt-5.5}} & \textbf{58} & \textbf{\code{gpt-5.5}} & \textbf{58} & \textbf{\code{gpt-5.6-luna}} \\
\textbf{56} & \textbf{\code{gemini-3.8-flash}} & \textbf{56} & \textbf{\code{gemini-3.8-flash}} & \textbf{58} & \textbf{\code{claude-opus-4-8}} \\
\textbf{56} & \textbf{\code{claude-fable-5-1}} & \textbf{56} & \textbf{\code{claude-fable-5-1}} & \textbf{52} & \textbf{\code{gemini-3.8-flash}} \\
\textbf{51} & \textbf{\code{gpt-5.6-sol}} & \textbf{52} & \textbf{\code{gpt-5.6-sol}} & \textbf{46} & \textbf{\code{claude-opus-5-5}} \\
\textbf{48} & \textbf{\code{claude-sonnet-5-5}} & \textbf{51} & \textbf{\code{claude-sonnet-5-5}} & \textbf{45} & \textbf{\code{gpt-5.5}} \\
\textbf{47} & \textbf{\code{grok-4.6}} & \textbf{43} & \textbf{\code{grok-4.6}} & \textbf{44} & \textbf{\code{claude-fable-5-1}} \\
\textbf{42} & \textbf{\code{deepseek-v4.1-flash}} & \textbf{41} & \textbf{\code{deepseek-v4.1-flash}} & \textbf{43} & \textbf{\code{claude-sonnet-5-5}} \\
25 & \code{claude-opus-4-8} & 23 & \code{claude-opus-4-8} & \textbf{39} & \textbf{\code{gpt-5.6-sol}} \\
\textbf{16} & \textbf{\code{gpt-5.6-luna}} & \textbf{21} & \textbf{\code{gpt-5.6-luna}} & \textbf{37} & \textbf{\code{deepseek-v4.1-flash}} \\
-- & \code{glm-5.3-flash} & -- & \code{glm-5.3-flash} & -- & \code{glm-5.3-flash} \\
-- & \code{gpt-5.4-mini} & -- & \code{gpt-5.4-mini} & -- & \code{gpt-5.4-mini} \\
\bottomrule
\end{tabular}}
\end{table}

\begin{table}[H]
\centering
\scriptsize
\setlength{\tabcolsep}{3pt}
\caption{\llms on C++, ranked by win probability (\%) in \latency, \energy, and \memory.}
\label{tab:bt-cpp}
\subfloat[\leetcodeComp\label{tab:bt-leetcode-99-perf-cpp}]{\begin{tabular}{@{}rl rl rl@{}}
\toprule
\multicolumn{2}{c}{\textbf{Latency}} & \multicolumn{2}{c}{\textbf{Energy}} & \multicolumn{2}{c}{\textbf{Peak memory}} \\
\cmidrule(lr){1-2}\cmidrule(lr){3-4}\cmidrule(l){5-6}
\textbf{67} & \textbf{\code{gpt-6-astra}} & \textbf{67} & \textbf{\code{gpt-6-astra}} & \textbf{52} & \textbf{\code{gpt-5.6-sol}} \\
\textbf{54} & \textbf{\code{gemini-3.8-flash}} & \textbf{54} & \textbf{\code{gemini-3.8-flash}} & \textbf{52} & \textbf{\code{gemini-3.8-flash}} \\
52 & \code{claude-fable-5-1} & 52 & \code{claude-opus-5-5} & \textbf{51} & \textbf{\code{claude-opus-4-8}} \\
52 & \code{gpt-5.4-mini} & 52 & \code{gpt-5.4-mini} & \textbf{51} & \textbf{\code{deepseek-v4.1-flash}} \\
51 & \code{gpt-5.6-sol} & 51 & \code{gpt-5.6-sol} & \textbf{51} & \textbf{\code{grok-4.6}} \\
51 & \code{claude-opus-5-5} & 51 & \code{claude-fable-5-1} & \textbf{51} & \textbf{\code{claude-opus-5-5}} \\
49 & \code{glm-5.3-flash} & 50 & \code{glm-5.3-flash} & \textbf{51} & \textbf{\code{gpt-5.4-mini}} \\
49 & \code{gpt-5.5} & 48 & \code{gpt-5.5} & \textbf{50} & \textbf{\code{claude-sonnet-5-5}} \\
47 & \code{deepseek-v4.1-flash} & 46 & \code{grok-4.6} & \textbf{49} & \textbf{\code{glm-5.3-flash}} \\
46 & \code{claude-sonnet-5-5} & 46 & \code{claude-sonnet-5-5} & \textbf{49} & \textbf{\code{gpt-5.5}} \\
46 & \code{grok-4.6} & 46 & \code{deepseek-v4.1-flash} & \textbf{48} & \textbf{\code{claude-fable-5-1}} \\
43 & \code{gpt-5.6-luna} & 44 & \code{claude-opus-4-8} & \textbf{48} & \textbf{\code{gpt-5.6-luna}} \\
43 & \code{claude-opus-4-8} & 44 & \code{gpt-5.6-luna} & \textbf{48} & \textbf{\code{gpt-6-astra}} \\
\bottomrule
\end{tabular}}\hfil
\subfloat[\clbgComp\label{tab:bt-clbg-10-cpp}]{\begin{tabular}{@{}rl rl rl@{}}
\toprule
\multicolumn{2}{c}{\textbf{Latency}} & \multicolumn{2}{c}{\textbf{Energy}} & \multicolumn{2}{c}{\textbf{Peak memory}} \\
\cmidrule(lr){1-2}\cmidrule(lr){3-4}\cmidrule(l){5-6}
\textbf{85} & \textbf{\code{gpt-6-astra}} & \textbf{83} & \textbf{\code{gpt-6-astra}} & \textbf{76} & \textbf{\code{deepseek-v4.1-flash}} \\
\textbf{72} & \textbf{\code{gpt-5.6-sol}} & \textbf{72} & \textbf{\code{gpt-5.6-sol}} & \textbf{66} & \textbf{\code{grok-4.6}} \\
\textbf{69} & \textbf{\code{gpt-5.5}} & \textbf{64} & \textbf{\code{gpt-5.5}} & \textbf{65} & \textbf{\code{gpt-5.6-luna}} \\
\textbf{60} & \textbf{\code{gemini-3.8-flash}} & \textbf{59} & \textbf{\code{gemini-3.8-flash}} & \textbf{56} & \textbf{\code{gpt-6-astra}} \\
\textbf{54} & \textbf{\code{glm-5.3-flash}} & \textbf{57} & \textbf{\code{glm-5.3-flash}} & \textbf{54} & \textbf{\code{claude-fable-5-1}} \\
\textbf{51} & \textbf{\code{claude-sonnet-5-5}} & 55 & \code{grok-4.6} & \textbf{50} & \textbf{\code{claude-sonnet-5-5}} \\
\textbf{50} & \textbf{\code{claude-fable-5-1}} & \textbf{52} & \textbf{\code{claude-fable-5-1}} & \textbf{45} & \textbf{\code{glm-5.3-flash}} \\
50 & \code{grok-4.6} & \textbf{50} & \textbf{\code{claude-sonnet-5-5}} & \textbf{43} & \textbf{\code{gpt-5.5}} \\
\textbf{49} & \textbf{\code{claude-opus-5-5}} & \textbf{49} & \textbf{\code{claude-opus-5-5}} & \textbf{43} & \textbf{\code{claude-opus-4-8}} \\
\textbf{37} & \textbf{\code{deepseek-v4.1-flash}} & \textbf{38} & \textbf{\code{deepseek-v4.1-flash}} & \textbf{42} & \textbf{\code{gpt-5.6-sol}} \\
32 & \code{claude-opus-4-8} & 32 & \code{claude-opus-4-8} & \textbf{38} & \textbf{\code{gemini-3.8-flash}} \\
\textbf{22} & \textbf{\code{gpt-5.6-luna}} & \textbf{23} & \textbf{\code{gpt-5.6-luna}} & \textbf{37} & \textbf{\code{claude-opus-5-5}} \\
18 & \code{gpt-5.4-mini} & 17 & \code{gpt-5.4-mini} & \textbf{36} & \textbf{\code{gpt-5.4-mini}} \\
\bottomrule
\end{tabular}}
\end{table}

\section{Problem Categories}
\label{sec:appendix-categories}

\Cref{tab:problem-categories} gives the number of \leetcodeComp problems in each \leetcode category that holds at least 9 of them, with one example problem each.

\begin{table}[H]
\centering
\footnotesize
\setlength{\tabcolsep}{3.5pt}
\caption{Category distribution of the 99~\leetcodeComp problems. Since a problem may belong to several categories, the counts sum to more than 99.}
\label{tab:problem-categories}
\begin{tabular}{@{}lrl@{}}
\toprule
Category & Problems & Example problem \\
\midrule
Array & 59 & \code{frog-jump} \\
String & 32 & \code{slowest-key} \\
Dynamic Programming & 24 & \code{2-keys-keyboard} \\
Hash Table & 22 & \code{word-ladder} \\
Math & 18 & \code{clumsy-factorial} \\
Sorting & 12 & \code{longest-string-chain} \\
Greedy & 11 & \code{di-string-match} \\
Bit Manipulation & 10 & \code{stickers-to-spell-word} \\
Backtracking & 10 & \code{ambiguous-coordinates} \\
Binary Search & 9 & \code{split-array-largest-sum} \\
Breadth-First Search & 9 & \code{cousins-in-binary-tree} \\
Depth-First Search & 9 & \code{tree-of-coprimes} \\
\bottomrule
\end{tabular}
\end{table}

\section{Calibrated \clbgComp Workloads}
\label{sec:appendix-clbg-workloads}

\Cref{tab:clbg-workloads} lists the workload of every \clbgComp benchmark on our host. Seven benchmarks take a command-line argument. The other three read a \bench{fasta} output from standard input, so the argument that generates this output sets their workload. With every workload, the C++ program runs for about one second.

\begin{table}[H]
\centering
\footnotesize
\caption{Calibrated workload of each \clbgComp benchmark on our host.}
\label{tab:clbg-workloads}
\begin{tabular}{@{}l r l@{}}
\toprule
\textbf{Benchmark} & \textbf{Workload} & \textbf{Meaning} \\
\midrule
\bench{binary-trees} & 18 & tree depth \\
\bench{fannkuch-redux} & 11 & permutation length \\
\bench{fasta} & 12,000,000 & output size \\
\bench{mandelbrot} & 3,504 & image side in pixels \\
\bench{n-body} & 15,000,000 & simulation steps \\
\bench{pidigits} & 10,000 & digits \\
\bench{spectral-norm} & 4,000 & matrix size \\
\midrule
\bench{k-nucleotide} & 2,500,000 & \bench{fasta} input, about 25~MB \\
\bench{regex-redux} & 500,000 & \bench{fasta} input, about 5~MB \\
\bench{reverse-complement} & 25,000,000 & \bench{fasta} input, about 254~MB \\
\bottomrule
\end{tabular}
\end{table}

\section{Check of the Reduced \leetcodeComp Workload}
\label{sec:appendix-workload-check}

\Cref{tab:workload-check} splits the check of the reduced \leetcodeComp workload (C6 in \Cref{sec:threats-construct}) by language. Both languages give about the same result. In \latency, the weighted average over the kept test cases and the average over all profiled test cases give the same duel outcome for 86.4\% of the duels in Python and 84.1\% in C++, and at most 1.7\% of the duels reverse. \Energy agrees less in both languages (85.5\% and 80.8\%), and at most 0.9\% of its duels reverse. Most of the other duels change between a tie and a win. The lowest values of~$\tau$ come from cells in which most solutions lie within the tie margin of each other. In the two cells with the lowest \latency~$\tau$, Python \code{minimum-number-of-increments-on-subarrays-to-form-a-target-array} ($\tau = 0.21$) and C++ \code{shift-2d-grid} ($\tau = 0.38$), 46 of the 78~pairs of solutions tie on all profiled test cases, so~$\tau$ there mostly orders differences within the noise. In these two cells, 88\% and 67\% of the duels keep their outcome, and two of the 78~duels reverse in each.

\begin{table}[H]
\centering
\footnotesize
\caption{Agreement between the two averages of the workload check, per language, on 10~random \leetcodeComp problems with the solutions of all \totalExecutedModels~\llms. Same: duels with the same outcome under the weighted average over the kept test cases and under the average over all profiled test cases, with the tie margins of our rankings. Reversed: duels in which a win becomes a loss. $\tau$: Kendall's~$\tau$ between the two orders of the \llms in a cell, median and minimum (in parentheses) over the cells.}
\label{tab:workload-check}
\begin{tabular}{@{}l l r r r r@{}}
\toprule
\textbf{Language} & \textbf{Metric} & \textbf{Duels} & \textbf{Same} & \textbf{Reversed} & $\boldsymbol{\tau}$ \\
\midrule
Python & Latency & 780 & 86.4\% & 1.4\% & 0.78 (0.21) \\
Python & Energy & 780 & 85.5\% & 0.4\% & 0.82 (0.31) \\
C++ & Latency & 780 & 84.1\% & 1.7\% & 0.79 (0.38) \\
C++ & Energy & 780 & 80.8\% & 0.9\% & 0.71 (0.33) \\
\midrule
Both & Latency & 1,560 & 85.3\% & 1.5\% & 0.79 (0.21) \\
Both & Energy & 1,560 & 83.1\% & 0.6\% & 0.79 (0.31) \\
\bottomrule
\end{tabular}
\end{table}

\section{Accept Rate by Difficulty}
\label{sec:appendix-difficulty}

\Cref{tab:difficulty-breakdown} breaks the accept rate down by \leetcode difficulty level for each \llm. \clbgComp has no difficulty levels, so this breakdown applies to \leetcodeComp only.

\begin{table}[H]
\centering
\footnotesize
\setlength{\tabcolsep}{6pt}
\caption{Accept rate by \leetcode difficulty on \leetcodeComp (percent), sorted by \passatone. The set holds 33~problems at each level, each measured in the \totalExecutedLanguages languages.}
\label{tab:difficulty-breakdown}
\begin{tabular}{@{}l rrr@{}}
\toprule
\textbf{\llm} & \textbf{Easy} & \textbf{Med.} & \textbf{Hard} \\
\midrule
\code{gpt-6-astra} & 100 & 100 & 99 \\
\code{claude-opus-5-5} & 100 & 100 & 97 \\
\code{gpt-5.6-sol} & 98 & 100 & 99 \\
\code{gpt-5.5} & 99 & 100 & 97 \\
\code{grok-4.6} & 100 & 99 & 97 \\
\code{claude-sonnet-5-5} & 99 & 100 & 96 \\
\code{claude-fable-5-1} & 100 & 100 & 94 \\
\code{deepseek-v4.1-flash} & 99 & 98 & 95 \\
\code{glm-5.3-flash} & 100 & 99 & 92 \\
\code{claude-opus-4-8} & 99 & 99 & 91 \\
\code{gemini-3.8-flash} & 99 & 98 & 90 \\
\code{gpt-5.6-luna} & 99 & 94 & 87 \\
\code{gpt-5.4-mini} & 99 & 94 & 73 \\
\bottomrule
\end{tabular}
\end{table}

The set holds 33~problems at each level. Eleven of the \totalExecutedModels \llms stay at or above $90\%$ at every level. The two weakest fall on the Hard problems, \gptfivemini from $99\%$ on Easy to $73\%$ on Hard and \gptluna from $99\%$ to $87\%$. The same two \llms also show blind spots in some problem categories. \gptfivemini accepts only $24\%$ of the three Number Theory problems and $69\%$ of the five Memoization problems, and \gptluna accepts $43\%$ of the Number Theory problems. The other \llms score higher in both categories. Number Theory has the lowest accept rate of all categories for every \llm (tied with Linked List for \gptsol), and its rate ranges from $24\%$ for \gptfivemini to $90\%$ for \gptastra. Each of these categories holds only three to five problems, so we report them as observations rather than established results.

\section{Cross-Language Correctness Consistency}
\label{sec:appendix-xlang}

\Cref{tab:xlang-consistency} gives the full breakdown behind the cross-language correctness analysis in the main text. Each \llm-problem pair is classified by how the accept or reject verdict varies across the \totalExecutedLanguages languages the \llm attempts.

\begin{table}[H]
\centering
\footnotesize
\caption{Cross-language correctness consistency on \leetcodeComp. Each \llm-problem pair is classified by its accept or reject verdict across the \totalExecutedLanguages languages (1,287 pairs attempted in at least two languages).}
\label{tab:xlang-consistency}
\begin{tabular}{@{}l r r@{}}
\toprule
\textbf{Verdict across languages} & \textbf{Pairs} & \textbf{Share} \\
\midrule
Accepted in every language & 1,148 & 89.2\% \\
Mixed (accepted in some, rejected in others) & 133 & 10.3\% \\
Rejected in every language & 6 & 0.5\% \\
\midrule
\textbf{Total} & 1,287 & 100\% \\
\bottomrule
\end{tabular}
\end{table}

The $6$ pairs rejected in every language are the only cases where an \llm solves a problem in none of the \totalExecutedLanguages languages. Five of them belong to \gptfivemini and one to \gptluna, and all fall on Hard problems, such as \code{abbreviating-the-product-of-a-range}. A mixed pair shows that the \llm can solve the problem but fails in some languages. Such a failure may come from a language-specific coding mistake, from a timeout in a slower language, or from sampling, because each cell holds one sample.

\section{Reference-Set Contributions}
\label{sec:appendix-baseline-contrib}

\Cref{tab:baseline-contrib} gives the full per-\llm, per-language breakdown of the reference set summarized in the main text. Each cell contributes its best correct solution for \latency, \energy, and \memory separately, so an \llm may win any metric of a cell independently. Each count is the number of these reference values the \llm supplies.

\begin{table}[H]
\centering
\caption{Reference solutions contributed by each \llm, by language. A cell is one problem in one language, and each cell contributes its best correct solution for \latency, \energy, and \memory separately, so an \llm may win any metric of a cell independently. Sorted by total.}
\label{tab:baseline-contrib}
\subfloat[\leetcodeComp\label{tab:baseline-contrib-leetcode-99-perf}]{\scriptsize
\setlength{\tabcolsep}{2.5pt}
\begin{tabular}{@{}l rrrrrrr r@{}}
\toprule
\textbf{\llm} & \textbf{C++} & \textbf{C\#} & \textbf{Java} & \textbf{JS} & \textbf{PHP} & \textbf{Py} & \textbf{TS} & \textbf{Total} \\
\midrule
\code{claude-fable-5-1} & \textbf{60} & 19 & 23 & 30 & 57 & 30 & 31 & \textbf{250} \\
\code{gpt-6-astra} & 31 & 30 & \textbf{39} & \textbf{46} & 17 & 43 & \textbf{41} & 247 \\
\code{claude-opus-5-5} & 22 & 31 & 22 & 30 & \textbf{67} & \textbf{44} & 26 & 242 \\
\code{claude-sonnet-5-5} & 17 & 28 & 24 & 22 & 35 & 39 & 25 & 190 \\
\code{gemini-3.8-flash} & 29 & 37 & 24 & 28 & 17 & 18 & 20 & 173 \\
\code{claude-opus-4-8} & 30 & \textbf{41} & 23 & 14 & 17 & 14 & 19 & 158 \\
\code{gpt-5.6-sol} & 21 & 13 & 24 & 16 & 15 & 24 & 30 & 143 \\
\code{glm-5.3-flash} & 23 & 21 & 15 & 27 & 18 & 8 & 21 & 133 \\
\code{gpt-5.6-luna} & 9 & 19 & 29 & 20 & 15 & 29 & 10 & 131 \\
\code{gpt-5.4-mini} & 14 & 14 & 25 & 20 & 8 & 9 & 19 & 109 \\
\code{gpt-5.5} & 10 & 24 & 14 & 19 & 11 & 13 & 14 & 105 \\
\code{grok-4.6} & 19 & 7 & 17 & 12 & 13 & 12 & 23 & 103 \\
\code{deepseek-v4.1-flash} & 12 & 13 & 18 & 13 & 7 & 14 & 18 & 95 \\
\midrule
\textbf{Total} & 297 & 297 & 297 & 297 & 297 & 297 & 297 & 2079 \\
\bottomrule
\end{tabular}
}\hfil
\subfloat[\clbgComp\label{tab:baseline-contrib-clbg-10}]{\scriptsize
\setlength{\tabcolsep}{2.5pt}
\begin{tabular}{@{}l rrrrrrr r@{}}
\toprule
\textbf{\llm} & \textbf{C++} & \textbf{C\#} & \textbf{Java} & \textbf{JS} & \textbf{PHP} & \textbf{Py} & \textbf{TS} & \textbf{Total} \\
\midrule
\code{gpt-6-astra} & \textbf{8} & \textbf{5} & 0 & 4 & 4 & \textbf{6} & \textbf{10} & \textbf{37} \\
\code{gpt-5.5} & 1 & 2 & 4 & \textbf{6} & 5 & 3 & 4 & 25 \\
\code{claude-fable-5-1} & 2 & 3 & \textbf{7} & 5 & 0 & 5 & 2 & 24 \\
\code{claude-opus-5-5} & 1 & \textbf{5} & 0 & 3 & \textbf{9} & 4 & 2 & 24 \\
\code{gemini-3.8-flash} & 3 & \textbf{5} & 2 & 3 & 1 & 3 & 0 & 17 \\
\code{grok-4.6} & 2 & 2 & 6 & 0 & 3 & 0 & 4 & 17 \\
\code{deepseek-v4.1-flash} & 2 & \textbf{5} & 3 & 0 & 1 & 0 & 2 & 13 \\
\code{gpt-5.6-sol} & 4 & 3 & 1 & 0 & 1 & 2 & 2 & 13 \\
\code{glm-5.3-flash} & 1 & 0 & 0 & 1 & 2 & 4 & 2 & 10 \\
\code{claude-opus-4-8} & 0 & 0 & 4 & 1 & 3 & 1 & 0 & 9 \\
\code{gpt-5.6-luna} & 3 & 0 & 1 & 4 & 0 & 1 & 0 & 9 \\
\code{claude-sonnet-5-5} & 3 & 0 & 1 & 1 & 1 & 0 & 1 & 7 \\
\code{gpt-5.4-mini} & 0 & 0 & 1 & 2 & 0 & 1 & 1 & 5 \\
\midrule
\textbf{Total} & 30 & 30 & 30 & 30 & 30 & 30 & 30 & 210 \\
\bottomrule
\end{tabular}
}
\end{table}

\section{Statistical Tests}
\label{sec:appendix-stat-tests}

\Cref{sec:ranking-method} defines the duels and the Bradley--Terry model, and \Cref{tab:stat-tests} lists the results of the tests that we describe here.

The \llm shuffle test checks whether the \llms differ at all in a metric. Its statistic is the standard deviation of the fitted strengths. If all \llms were equally efficient, the solutions in a cell would be interchangeable. We therefore shuffle, inside each cell, which of the \llms that solved the cell produced which solution, and we refit the model 1,000~times. The $p$-value is the share of shuffles whose strengths spread at least as much as the observed ones.

To find which pairs differ, we draw 10,000 bootstrap samples~\cite{Efron1979Bootstrap} and refit the model on each sample. For each of the $13 \times 12 / 2 = 78$~pairs, the standard deviation of the strength gap over the samples gives a standard error, and a two-sided $z$~test of the observed gap gives a $p$-value. Holm's procedure~\cite{holm1979} corrects the 78~$p$-values of a metric, so that the chance of reporting even one false difference stays below 5\%. The cells of one problem in different languages are not independent, so on both competitions the bootstrap resamples whole problems. \Cref{tab:stat-tests} reports both resamplings for both competitions. On \leetcodeComp, resampling cells instead of problems finds more significant pairs (53 instead of 42 in \latency and 52 instead of 39 in \energy) and drops \geminiflash from the tied leaders in \energy. On \clbgComp, resampling cells finds 33 instead of 21 pairs in \latency, 23 instead of 17 in \energy, and 6 instead of 5 in \memory (I7 in \Cref{sec:threats-internal}).

The language shuffle test checks whether the language changes the order of the \llms. It compares the per-language rankings with Kendall's~$\tau$. Two languages would disagree somewhat even if the language did not matter, because each per-language ranking rests on fewer duels. We therefore shuffle, inside each problem, which language each measurement belongs to, recompute the mean~$\tau$ 1,000~times, and report as~$p$ the share of shuffles whose~$\tau$ is at most the observed one. The mean~$\tau$ over the shuffles is the chance level, and a $p$ below $0.05$ means that two languages agree less than chance allows.

The tie margin barely changes the rankings. With no margin, so that every difference decides a duel, the \latency and \energy rankings agree with ours at Kendall's~$\tau = 0.97$ on both competitions. Only the \memory rankings change ($\tau = 0.69$ on \leetcodeComp and $0.62$ on \clbgComp), because many \memory differences lie within the noise.

A ranking by the geometric mean of the ratios to the reference set, taken over the cells that each \llm solved, would favor the \llms that fail hard problems. By that measure, \gptfivemini ranks second in \latency on \leetcodeComp, because it fails 80~cells, 62 of them on Hard problems. By win probability, it ranks ninth, because each of its duels compares it with another \llm on a cell that both solve.

\begin{table}[H]
\centering
\footnotesize
\setlength{\tabcolsep}{4pt}
\caption{Statistical tests of the Bradley--Terry rankings. Duels: median number of duels per \llm pair. $p$: \llm shuffle test of whether the \llms differ at all. Sig.\ pairs: pairs that differ after Holm's correction, when the bootstrap resamples cells or whole problems. Leaders: \llms that no other \llm beats significantly, from the problem bootstrap on both competitions. Language: mean Kendall's~$\tau$ between the per-language rankings, its chance level, and the $p$ of the language shuffle test.}
\label{tab:stat-tests}
\begin{tabular}{@{}l r r r r r r r r@{}}
\toprule
 & & & \multicolumn{2}{c}{\textbf{Sig.\ pairs}} & & \multicolumn{3}{c}{\textbf{Language}} \\
\cmidrule(lr){4-5}\cmidrule(l){7-9}
\textbf{Metric} & \textbf{Duels} & $\boldsymbol{p}$ & \textbf{Cells} & \textbf{Problems} & \textbf{Leaders} & $\boldsymbol{\tau}$ & \textbf{Chance} & $\boldsymbol{p}$ \\
\midrule
\multicolumn{9}{@{}l}{\leetcodeComp} \\
Latency & 661 & $<0.001$ & 53/78 & 42/78 & 2 & $0.50$ & $0.72$ & $<0.001$ \\
Energy & 661 & $<0.001$ & 52/78 & 39/78 & 3 & $0.48$ & $0.71$ & $<0.001$ \\
Peak memory & 661 & $0.092$ & 0/78 & 0/78 & 13 & $0.06$ & $0.06$ & $0.602$ \\
\midrule
\multicolumn{9}{@{}l}{\clbgComp} \\
Latency & 51 & $<0.001$ & 33/78 & 21/78 & 8 & $0.55$ & $0.54$ & $0.518$ \\
Energy & 50 & $<0.001$ & 23/78 & 17/78 & 8 & $0.46$ & $0.49$ & $0.248$ \\
Peak memory & 51 & $0.019$ & 6/78 & 5/78 & 8 & $0.09$ & $0.12$ & $0.138$ \\
\bottomrule
\end{tabular}
\end{table}

\section{Additional Analyses}
\label{sec:appendix-additional}

Energy consumption diverges from \latency in magnitude on \clbgComp but not on \leetcodeComp. Here, a ratio is the ratio to the reference set (\Cref{sec:reference-set}), and we take its geometric mean over the cells that an \llm solves. On \leetcodeComp, the two ratios stay within $2\%$ for every \llm. On \clbgComp, every \llm's \energy ratio is below its \latency ratio ($13$ of $13$ \llms, sign test $p < 0.001$, the probability of $13$ such \llms out of $13$ if a higher and a lower \energy ratio were equally likely), by $7\%$ on average and by up to $15\%$ for \opusfour (\latency $3.91$, \energy $3.32$), so a slow whole-program solution pays a smaller \energy penalty than its time penalty. This may be because the fastest solution of a \clbgComp cell draws more package power than the median of the other solutions in 49 of the 70~cells, for example when it uses both cores of the host.

For \memory, the cost per attempted solution relates to the win probability on neither competition, neither in USD ($\rho = +0.34$ on \leetcodeComp and $-0.19$ on \clbgComp) nor in tokens ($\rho = -0.09$ and $+0.17$), with $p \ge 0.25$ for all four.

The cost of a correct solution also depends on the language. For each \llm, we order the seven languages by its cost per correct solution, and a Friedman test~\cite{friedman1937} checks whether the \llms agree on this order. They agree on both competitions, in USD ($\chi^2_F = 46.1$ on \leetcodeComp and $21.2$ on \clbgComp) and in tokens ($\chi^2_F = 45.4$ and $20.5$), with $p \le 0.003$ for all four. Python is the cheapest language on \leetcodeComp for 10~\llms and PHP the most expensive for 12, in both USD and tokens. On \clbgComp, JavaScript is the cheapest for 7~\llms and C++ the most expensive for 7, again in both units.

We also check whether the \llms generate better or cheaper code in more popular languages, because popular languages may dominate the training data of the \llms. We rank the languages by their use on GitHub in 2025~\cite{GitHub2025Octoverse}. For each language, we compute the median over the \llms of the ratio to the reference set in the language for each performance metric, the slowdown of the language against the fastest language on the same problems, the mean pass rate, and the median cost per correct solution in USD and in tokens. With seven languages, Spearman's~$\rho$ reaches significance at the $0.05$ level only beyond $\pm 0.79$. We find no evidence that popularity relates to any of these measures on either competition ($|\rho| \le 0.75$, $p \ge 0.05$). The strongest trends are that less popular languages cost more per correct solution on \leetcodeComp ($\rho = +0.75$, $p = 0.052$) and have a lower pass rate there ($\rho = -0.75$, $p = 0.052$). The cost trend is weaker on \clbgComp ($\rho = +0.54$, $p = 0.22$).

\twocolumn
\bibliographystyle{IEEEtranN}
\bibliography{bibliography}

\end{document}